# How metal films de-wet substrates – identifying the kinetic pathways and energetic driving forces

Kevin F. McCarty[1], John C. Hamilton[1], Yu Sato[2], Angela Saá[3], Roland Stumpf[4], Juan de la Figuera[3,5], Konrad Thürmer[1], Frank Jones[1], Andreas K. Schmid[2], A. Alec Talin[1], and Norman C. Bartelt[1]

[1]Sandia National Laboratories, Livermore, CA
[2]Lawrence Berkeley National Laboratory, Berkeley, CA
[3]Universidad Autónoma de Madrid, Madrid Spain
[4]Sandia National Laboratories, Albuquerque, NM
[5]Instituto de Química-Física "Rocasolano", CSIC, Madrid Spain

mccarty@sandia.gov

**Abstract**. We study how single-crystal chromium films of uniform thickness on W(110) substrates are converted to arrays of three-dimensional (3D) Cr islands during annealing. We use low-energy electron microscopy (LEEM) to directly observe a kinetic pathway that produces trenches that expose the wetting layer. Adjacent film steps move simultaneously uphill and downhill relative to the staircase of atomic steps on the substrate. This step motion thickens the film regions where steps advance. Where film steps retract, the film thins, eventually exposing the stable wetting layer. Since our analysis shows that thick Cr films have a lattice constant close to bulk Cr, we propose that surface and interface stress provide a possible driving force for the observed morphological instability. Atomistic simulations and analytic elastic models show that surface and interface stress can cause a dependence of film energy on thickness that leads to an instability to simultaneous thinning and thickening. We observe that de-wetting is also initiated at bunches of substrate steps in two other systems, Ag/W(110) and Ag/Ru(0001). We additionally describe how Cr films are converted into patterns of unidirectional stripes as the trenches that expose the wetting layer lengthen along the W[001] direction. Finally, we observe how 3D Cr islands form directly during film growth at elevated temperature. The Cr mesas (wedges) form as Cr film steps advance down the staircase of substrate steps, another example of the critical role that substrate steps play in 3D island formation.

# 1. Introduction

Many film/substrate systems are unstable if the film uniformly covers the substrate. That is, the thermodynamically preferred configuration is not a flat film but separate, three-dimensional (3D) islands [1]. The area between the 3D islands exposes either the bare substrate or a thin, uniform wetting layer of film [2]. In the latter case (i.e., Stranski-Krastanov systems), the wetting layer is thermodynamically stable. These systems form 3D islands because the surface energy of the substrate is less than the sum of the film/substrate interface energy and the film's surface energy [2-4]. However, even in systems that tend to de-wet and form 3D islands, low-temperature deposition can yield relatively flat films. This approach for producing uniform films, which many technologies depend upon [5], is successful if the kinetic processes that de-wet the film are slow. Since the atomic processes that convert a flat film into 3D islands are poorly known, predicting the time needed for film de-wetting is not now possible. From the opposite point of view, fast de-wetting is required if 3D structures are the desired end state. Well-defined 3D structures such as dots and wires on surfaces have great technological potential [6, 7]. Hence, research has been devoted to determining whether de-wetting is a viable route to synthesizing self-organized nanostructures [8-11]. But again this research is stymied because the detailed atomic mechanisms for 3D island formation remain unknown.

For continuous films to de-wet, they must thin sufficiently in local regions to expose either the bare substrate or the thin wetting layer. In other regions the film must thicken so 3D islands can form. Unlike liquid films, thinning and thickening crystalline films below the roughening temperature involves local removal and addition of discrete atomic layers. Film defects are generally thought to trigger de-wetting, particularly by exposing the substrate (or wetting layer). In polycrystalline films, for example, grain-boundary grooving can provide a low-energy-barrier pathway to thin the film to expose either the bare substrate or the wetting layer [12-14]. Similarly, impurities can facilitate thinning. For example, film layers can retract from impurity particles, thus exposing the substrate (wetting layer) next to the particle [15]. While clean, single-crystal films cannot use the grain-boundary and impurity pathways of de-wetting, threading, screw-type dislocations can enable de-wetting.

But what if the single-crystal film contains no threading screw dislocations? Both the thinning and thickening processes in defect-free (single-crystal) films that are needed to expose the substrate (wetting layer) would seem, at first glance, to require nucleating pits in film layers and nucleating new film layers, respectively (see Fig. 1). Yet the energy cost associated with the step bounding a new layer causes an energy barrier for new-layer nucleation [16, 17] that can prohibit this thickening mechanism. De-wetting-induced nucleation can only be observed if a sufficiently strong driving force for de-wetting and a low step-edge energy are present [18]. For the systems discussed here, that is, metal-on-metal films at moderate temperatures, the material itself cannot generate a sufficient supersaturation of mobile adspecies to nucleate new layers [a]. A similar energy barrier exists to nucleating an atomic pit in a facet. Thus, hole nucleation is similarly improbable unless film atoms are being removed from the surface by sublimation or sputtering, for example. Since the simple-minded pathways of Fig. 1 are unlikely, a key question is: what are the microscopic processes that thin and thicken defect-free films during de-wetting?

Here we reveal a process by which a continuous film simultaneously thins to expose the wetting layer in local regions and thickens in adjacent regions, eventually forming 3D islands. We study the system of single-crystal Cr films on a W(110) substrate [19], a system free of threading dislocations. Cr films grown near room temperature are rough. When annealed, the film surface first smoothes to uniformly cover the substrate. With further annealing, the films de-wet. Because of this system's large anisotropy, the trenches that expose the stable wetting layer form only along one crystallographic direction, the W[001] direction. The Cr islands formed by de-wetting are arrays of 3D, unidirectional stripes. Fe also forms stripes when de-wetting W(110) [20-22] [b]. As we will show, this directionality allows observing the nucleation of individual trenches with sufficient temporal and spatial resolution so the nucleation event can be described by the motion of individual atomic film steps.

We find that atomic steps, both on the film's surface and at the film/substrate interface, allow a kinetic pathway that avoids the nucleation of holes in film layers to expose the wetting layer or the nucleation of new film layers to thicken the film. This finding shows that the continuum approaches commonly used to model thin-film morphological instabilities does not always adequately describe the essential elements of the de-wetting process [32]. The Cr films we study are largely unstrained by the substrate. (They have a lattice constant indistinguishable by low-energy electron diffraction from bulk Cr.) Yet the films still thin and thicken in adjacent regions even before the wetting layer is exposed. We use atomistic simulations and analytic models to analyze the energetics of films whose lattice constants are not strained ("clamped") to match the substrate's lattice parameter. Because of a driving force that arises from surface and interface stress, we find that these "unclamped" flat films are unstable relative to thickness undulation. We also find that the morphology of the 3D Cr islands is dominated by the kinetic pathways of thinning and thickening. That is, even after extensive annealing, the Cr islands remain long stripes, far from their equilibrium shape, because the stripes can thicken and narrow efficiently compared to compact, equilibrated islands. In addition, we discuss the relevance of the mechanism we observe to de-wetting of crystalline films on amorphous substrates. Finally, we examine how 3D Cr islands form directly during deposition at elevated temperature. We find that substrate steps also facilitate 3D island formation during film growth, as during annealing.

---

[a] During film growth, the deposition flux can provide sufficient supersaturation for new-layer nucleation.

[b] Islands elongated along the [001] direction commonly form during heteroepitaxial or homoepitaxial growth at elevated temperature of bcc metals on bcc (110) surfaces [23-31].

This manuscript has the following organization. In the next section (2), we discuss experimental methods. In section 3 we present observations that reveal how a Cr film simultaneously thins and thickens to expose the wetting layer. In section 4 the energetics of the observed film thinning and thickening are analyzed. We show how surface and interface stress can cause the film energy to vary with thickness in a manner consistent with the observed behavior. In section 5 we show that de-wetting in two other systems (Ag on Ru(0001) and Ag on W(110)) also is initiated at substrate steps. We then discuss the generality of our proposed mechanism. In section 6 we provide considerable detail about how Cr forms 3D stripe patterns during de-wetting. In section 7 we describe how film de-wetting occurs during growth at elevated temperature. This section ends with a discussion of the physical origins of anisotropy in the Cr/W(110) system.

## 2. Methods

### 2.1. Experimental methods

Most experiments were performed in an ultrahigh-vacuum-based (UHV) low-energy electron microscope (LEEM). The substrate was a disk-shaped single crystal of tungsten cut and polished using standard metallographic procedures to give a (110) surface about 10 mm in diameter. In the LEEM, this surface was cleaned by exposing to $2 \times 10^{-8}$ T of $O_2$ at 1000°C followed by flashing several times to 1650°C. Cr and Ag were evaporated from high-purity metals (99.996% for Cr and 99.999% for Ag) held in Mo crucibles heated with electron beams. The Cr deposition rate was measured before and after each de-wetting experiment by observing the time needed to deposit the first complete Cr layer on W(110) at 455°C, a temperature sufficiently high to give perfect step-flow film growth [33, 34]. Films between 10-40 monolayer (ML) thick for the de-wetting experiments were deposited at or somewhat below room temperature. To observe the de-wetting mechanism, we imaged the films while annealing at temperatures well below that where sublimation begins [19]. For the conditions (6 eV electrons) used to image de-wetting, the 1 ML Cr wetting layer appears dark and thicker Cr appears bright.

The great majority of the substrate surface analyzed by LEEM is covered by a fairly uniform array of monatomic steps. While the miscut of the crystal's surface normal away from the perfect [1$\bar{1}$0] direction varied locally, there is a well-defined average miscut. On average the staircase of substrate steps is at an angle of roughly 45° to the W[001] direction. The local uphill and downhill directions along the substrate staircase were determined by monitoring with LEEM the directions that Cr film steps advanced during deposition under conditions of step-flow growth. Steep step bunches, as well as some pits, such as in Fig. 2, occur at very low density. The stripe-shaped Cr islands and trenches that expose the wetting layer were determined to lie along the W[001] direction by calibrating the image rotation between diffraction and real-space imaging. For films thinner than about 6 layers, the local film thickness was determined in two ways -- either directly during step-flow growth by tracking film step edges, or by measuring how the electron reflectivity varied as a function of electron energy (i.e., the quantum-size effect [35]). Atomic-force microscopy (AFM) was performed on a Cr stripe array prepared in the LEEM and removed to air for analysis.

Some de-wetting experiments were performed in a 2[nd] UHV chamber on a different W(110) crystal using scanning tunneling microscopy (STM) as the principal diagnostic tool (see section 6.4). The crystal was prepared and Cr films grown using procedures similar to those described above. The crystallographic directions of this crystal were determined using STM to image the directions of the atomic rows in the (2×1) structure of oxygen adsorbed on W(110) [36].

## 3. Observing the kinetic pathway of de-wetting

### 3.1. Cooperative motion of film steps exposes the wetting layer

The LEEM images in Fig. 2 show the initial stages of a Cr film de-wetting a W(110) substrate. After room-temperature growth, the film surface was very rough, as evidenced by the almost complete lack of specularly reflected electrons in LEEM. During annealing the film smoothes considerably, and becomes almost conformal with the substrate. That is, the film steps lie nearly on top of substrate steps. But this conformal film is unstable. With further annealing the film begins to de-wet by forming trenches that penetrate down to the stable Cr wetting layer (1 ML of Cr). These trenches, the dark, horizontal bands in Fig. 2 (other images are aligned along other directions) form where the substrate has a high density of atomic steps. That is, the trenches in Fig. 2 form only at the bunch of substrate steps that bound the circular pit in the substrate. This observation reveals the first clue about the de-wetting mechanism -- de-wetting is initiated at regions of high substrate step density.

Further insight comes from high-resolution imaging of individual trenches forming, as shown in Fig. 3. The images contain a finger-shaped region bounded by dark bands, which are bunches of atomic Cr steps. Real-time observations show that the finger formed by Cr steps advancing down its length, from the top to the tip. This direction is also roughly the direction along which the substrate steps descend. The process of film steps advancing down the substrate steps thickens the film, as schematically illustrated in Fig. 3. Thus, the film finger is thicker than the surrounding film regions. This finger draws material from the adjacent regions, which, consequently, get *thinner*. This conclusion is documented by comparing the location of the Cr step bunch marked by the arrow in the 180-s image to its location in the 1-s image – the step bunch has moved up, which is roughly up the staircase of substrate steps. This direction of Cr step motion thins the film, in contrast to the step motion in the downhill direction, which thickens the fingers. A trench that exposes the 1 ML wetting layer nucleates in the 194-s image. Inspection of the still images in Fig. 3 and the accompanying video clearly show that the trench nucleates where the Cr steps are moving uphill (i.e., at the arrow in the 180-s image). Figure 4 and its accompanying video give another example of the Cr step motion that exposes the wetting layer as mass is transferred from regions being thinned by film steps moving uphill to regions being thickened by film steps moving downhill.

## 3.2. Mechanism of de-wetting

The above observations establish the de-wetting mechanism, which is shown schematically in Fig. 5 [c]. An instability develops when a film step randomly advances across a descending substrate step. The mass transferred to the advancing film step is supplied by the retraction of an adjacent film-step section. As discussed in section 4, the initially flat film is unstable to undulations consisting of film regions thinner and thicker than the average (initial) thickness. Thus, the undulations grow in size as more surface area thins by film steps retracting up the staircase of substrate steps. This film material is transferred to regions that thicken by steps advancing down the staircase of substrate steps. Eventually a retracting step exposes the wetting layer. De-wetting proceeds as more of the wetting layer is exposed by retracting film steps. This mass flows towards and advances steps of the thick regions, which develop into 3D Cr stripes, as discussed in detail in section 6. To summarize, the kinetic pathway that exposes the wetting layer proceeds by atomic film steps moving both up and down relative to the staircase of substrate steps. This mechanism does not involve the nucleation of new steps. Thus, the energetically costly pathways of Fig. 1 are avoided. However, the mechanism of Fig. 5 increases the length of existing film steps, as discussed in more detail in section 6.4.

## 3.3. Cr films are unstrained by the substrate

A key issue to understanding the energetic driving forces of de-wetting is to determine whether the substrate imposes a strain on the Cr films. Only films 1 to 3 layers thick are pseudomorphic and, thus, highly strained [d] [34]. Thicker films have interfacial dislocations that largely relax the strain. As Fig. 6 shows, LEED from de-wetted films (i.e., with both Cr stripes and the pseudomorphic wetting layer) have two sets of well-separated diffraction spots – those from the W substrate plus pseudomorphic wetting layer, and those from the thick film portion, i.e., the 3D Cr stripes. Using the diffraction spots from the pseudomorphic wetting layer as an internal calibration, we find that the in-plane lattice parameter of thick Cr is 2.84±0.06 Å, based on measurements from three different films. For comparison, the lattice parameters of bulk Cr and W are 2.88 and 3.165 Å, respectively [37]. Thus, the de-wetted Cr films have essentially the same lattice parameter as bulk Cr. Rotenberg and coworkers used LEED to shown that the in-plane lattice parameter of annealed, de-wetted films is the same (to high precision) as those of annealed, uniform films [33]. This observation, when combined with the above analysis, establishes that annealed, uniform Cr films are stain-free at the accuracy level of LEED. Additional support for this conclusion comes from a detailed structural study, which showed that the out-of-plane relaxations of the topmost planes in annealed, 15-ML-thick Cr films also agree within error with the values expected for a bulk Cr(110) surface [34] e. A dislocation network that lies close to the substrate/film interface allows the Cr film to relax in-plane to the bulk Cr lattice parameter. We have not determined exactly where the dislocations lie. Most likely they lie in the plane between the first film layer and substrate, or within the first few film layers [39, 40] f.

---

[c] For simplicity, the wetting layer and substrate are both colored red.

[d] Diffraction from films between 4 and roughly 10 ML thick contain superstructure spots, which result from the substrate and film having different lattice constants. That is, these films are non-pseudomorphic. This point is discussed further in section 7.2. Thicker film do not have superstructure diffraction spots, only the simple tetragonal diffraction pattern of Cr(110) [34] because the electrons do not penetrate to and diffract from the substrate. We also note that 3 ML of Cr is on the border of the pseudomorphic/non-pseudomorphic (dislocated) transition. In a minority of observations using selected area diffraction, the (00), and first-order diffraction spots of 3 ML Cr were weakly split ('streaked') along the [001] direction [34]. In addition, the 3 ML regions contain obvious linear defects in LEEM images. The defects can be weakly seen in Fig. 25 even though the imaging conditions were optimized for thickness contrast.

So the cooperative motion of film steps (Fig. 5) we observe occurs even though the Cr films are not fixed to the lattice parameter of the substrate. Thus, even in systems where the substrate is not imposing a measurable strain on the film, the film can be unstable relative to local thinning and thickening. We next discuss a possible energetic driving forces for unclamped films to simultaneously thicken and thin and eventually de-wet.

## 4. Energetics of de-wetting

### 4.1. Energy considerations in de-wetting

A key observation obtained from our real-time LEEM measurements is that Cr is transferred from some local regions, thinning them, to adjacent regions, thickening them. The atomic model in Fig. 7 illustrates transferring mass from a terrace's descending step edge to its ascending step edge. In the example, the mass flow reduces the area of the 6-layer film by making more 5- and 7-layer film. While such mass flow can ultimately expose the wetting layer, we experimentally observe the mass flow *before* the wetting layer is exposed (Figs. 2-4). Thus, one cannot evoke the de-wetting driving

---

e We note that the Cr films are annealed at high temperature, typically above 500°C, before de-wetting occurs. During this annealing, the film greatly smoothes, showing that significant mass diffusion occurs. This annealing can relieve any stress present in the as-deposited film [38].

f Popescu et al. [38] have proposed a structural model for the dislocation network of Fe films on W(110) based on x-ray diffraction. However, the precise location of the network relative to the substrate was not determined because of the ambiguity about whether pseudomorphic Fe layers exist.

force (i.e., differences in interfacial energies, see section 6.1) to explain the mass flow that simultaneously thickens and thins the film. That is, the system cannot reduce the area of the film/substrate interface until the substrate (or wetting layer) is exposed.

Thin films strained so that their lattice constants match the substrate's are known to be unstable to mass flows such as Fig. 7. In these systems, elastic relaxations can be larger at higher step edges, which is the physical basis for the well-known Asaro-Tiller-Grinfeld (ATG) [41] instability (see section 5.3). Thus, a flat film can reduce its strain energy by becoming undulated. But as discussed in section 3.3, the Cr films are not measurably strained by the substrate and the Cr is relaxed to the condition of bulk Cr. But we observe the Cr films to be unstable even though the ATG theory seems to be inappropriate for the system. So why should the concerted step motion (which increases total step length and surface energy) occur?

In this section, we show how *surface stress* can provide a driving force that causes a morphological instability even in films that are not fixed to the lattice constant of the substrate. Surface stress arises because atoms at surfaces are under coordinated, resulting in a tendency to reduce the bond lengths to their neighbors [41-43]. This causes the well-known relaxation of surface atoms, usually towards the underlying layers. Surface atoms also have a tendency to reduce their bond lengths in the surface plane, resulting in surface stress. Thus, surface stress is an intrinsic property and occurs even in films not strained by substrates. For films only a few atomic layers thick, the surface atoms actually reduce their in-plane spacing. Thus, the interior ("bulk") layers of very thin films are also compressed in-plane relative to the bulk lattice spacing [44]. Surface stress is reduced at the cost of straining the "bulk".

In the remainder of this section, we explore the potential role of surface stress in the morphological stability of thin films. In section 4.2 we use atomistic simulations to show that an unclamped, flat film is unstable relative to undulations. We then (section 4.3) present a simple spring model that describes how surface stress causes the energy of an unclamped film to vary with thickness. This dependence of the film energy on thickness provides a driving force to roughen unclamped films. In section 4.4, atomistic simulations are used to directly calculate the film energy versus thickness. Finally, in section 4.5, a detailed elasticity model is used to show that this energy dependence on thickness arises from surface stress.

### 4.2. Atomistic simulation of simultaneously thinning and thickening a stepped slab

Again we note that dislocations that lie within the first few Cr layers relax the strain so that thicker layers are not clamped at the lattice parameter of the substrate. Thus, the Cr film can be considered to be a free-standing slab with upper and lower surfaces. The upper surface represents the film/vacuum interface and contains the usual surface stress. The lower surface of the film -- the heterophase interface between the Cr and W -- is also subject to an interfacial stress. That is, the interfacial energy will vary continuously with film strain, at least for small strains, because of a continuous change in the misfit dislocation density. It is possible that this interfacial stress is large compared to the stress of the free surface. For simplicity, we take the free upper surface and the interface with the substrate to have the same properties in the atomistic simulations and the analytical models described below [g].

We first present atomistic simulations using the embedded-atom method (EAM) [45] to determine how the energy of thin metal slabs change with slab thickness. Modeling Cr using semi-empirical potentials requires major changes in the EAM potentials to introduce the non-central forces that are needed to reproduce phenomena such as the negative Cauchy pressure of Cr [46, 47]. Thus, in order to illustrate the relevant phenomena, we consider a free-standing slab with two (111) surfaces of Cu, which is accurately modeled using the EAM potential of Foiles, Baskes, and Daw [45]. We emphasize that our conclusions are general, at least to metals, and are not specific to Cu(111) nor fcc metals.

Figure 7 shows two Cu(111) slabs, unclamped in the directions parallel to the surfaces, with a stepped upper surface and a bottom surface constrained to be flat. Both slabs have periodic boundary conditions in the two in-plane directions. Since the experimental films are unclamped, we allow the periodic boundary lengths to change during the relaxation to the minimum energy state. The slab bottom represents the interface with the substrate (or film layer that contains the dislocations). The two slabs contain the same number of atoms but differ in their spatial distribution. The lower slab is derived from the upper slab by taking two rows of atoms from the edges of the 6-atom-thick regions and placing them at the edges of the 7-atom-thick region. This change gives the lower slab greater areas of thinner (5-atom-thick) and thicker (7-atom-thick) film compared to the upper slab. After carefully relaxing the slabs, we find that the lower slab is 0.1 meV per moved atom lower in energy than the upper slab. (We show in section 4.5 that essentially this same energy difference results from an elasticity model of surface stress.) Thus, the film can lower its energy by transferring atoms to thin some regions while thickening other regions.

### 4.3 Simple elasticity model of de-wetting driven by surface stress

In this section we discuss the driving force for the instability revealed in the EAM simulations. A number of computations have shown that the topmost atomic layer, the "surface monolayer" of free-standing metallic nano-sheets, are in tensile stress. In contrast, interior ("bulk") layers are in compressive stress [44, 48, 49]. At equilibrium, the in-plane stresses of the surface monolayer and bulk layers must be equal but opposite. As the film becomes thinner, the in-plane lattice constant becomes smaller because there is less and less bulk material to be compressed by the surface.[h] In this and the next section, we examine the effect of the bulk and surface stresses on the film energy using analytical elastic models. Using these models, we derive the dependence of film energy on thickness and show how surface stress can provide a driving force for simultaneously thinning and thickening a film.

---

[g] A more detailed treatment would let the free surface and the Cr surface in contact with the dislocated Cr layers have different surface (interfacial) stresses. However, this detail will not change our qualitative conclusion that surface and interface stress causes thinner layers to be more stable than thicker layers.

We define $E(h)$ as the energy per surface film atom, relative to bulk Cr, of a film with $h$ atomic layers. For mass to flow from a film terrace of intermediate thickness, $h_i$ ($h_i = 6$ in Fig. 7), to terraces one layer thinner and thicker ($h_{i-1}$ and $h_{i+1}$, respectively), we must have:

$$E(h_{i-1}) + E(h_{i+1}) - 2E(h_i) < 0, \text{ or:}$$

$$\frac{d^2E}{dh^2} < 0 \qquad\qquad (1).$$

Thus, the shape of $E(h)$ should be concave down.

The simplest model that can illustrate this effect is a one-dimensional spring model, as proposed by Müller and Saúl [41]. In this model, the bonds within the slab surfaces are represented by springs having length, $L$, spring constant, $k_s$, and equilibrium length, $a$. For $N$ surface atoms the energy of a single surface layer is:

$$E_{surf} = \frac{Nk_s}{2}(L-a)^2.$$

We call this quantity the "surface-monolayer energy".

Similarly, the bonds within the bulk layers are represented by springs having length, $L$, spring constant, $k_b$, but different length $b$. Assuming a pseudomorphic structure with $N$ atoms in a bulk monolayer, the energy of a single bulk layer is:

$$E_{bulk} = \frac{Nk_b}{2}(L-b)^2.$$

We call this quantity the "bulk-monolayer energy".

Now consider a slab consisting of $h$ monolayers, which will have 2 surface layers and $h$-2 bulk layers. The total energy per surface atom is:

$$E(L) = \frac{1}{2}\Big[2k_s(L-a)^2 + (h-2)k_b(L-b)^2\Big] \qquad\qquad (2).$$

Differentiating and finding the minimum energy state by solving for $dE/dL = 0$ gives the bond length of the relaxed slab as:

---

[h] In principle the variation of lattice constant with film thickness could be experimentally determined. However, the variation is anticipated to be quite small. For example, EAM simulations predict that the strain of Cu(111) slabs changes by about 0.002 for a thickness change from 6 to 10 atomic layers [50]. Because changes of this size are smaller than the precision of typical LEED, we have not attempted to measure the Cr lattice constant as a function of film thickness.

$$L = \frac{b(h-2)k_b + 2ak_s}{(h-2)k_b + 2k_s}.$$

Substituting this equilibrium bond length into Eqn. 2, we find:

$$E(h) = \frac{(a-b)^2(h-2)k_b k_s}{(h-2)k_b + 2k_s}$$

and, then,

$$\frac{d^2E}{dh^2} = -\frac{4(a-b)^2 k_b^2 k_s^2}{[(h-2)k_b + 2k_s]^3}.$$

This second derivative is negative. Thus, according to the criteria of Eqn. 1, the film can lower its energy by thinning some film regions and thickening others. We conclude that surface stress alone provides a driving force to undulate a flat film, i.e., an unclamped film can lower energy by separating into thinner and thicker regions.

### 4.4  Atomistic simulation of film energy vs. thickness

We have also used EAM calculations of thin, uniformly-thick slabs to calculate directly the function $E(h)$. For these calculations, Cu slabs with free surfaces in the [111] direction and with periodic boundary conditions in the $[1\bar{1}0]$ and $[11\bar{2}]$ directions were relaxed to their minimum energy configurations. This relaxation also allowed the periodic boundary lengths in the in-plane directions to vary, thereby adjusting the in-plane lattice constants so the tensile stress of the Cu(111) surfaces balanced the compressive stress of the Cu bulk. These calculations were repeated for thicknesses $h =$ 2-11. The filled circles in Fig. 8 plot $E(h)$ with units eV/atom and with the energy zero being that of an unstrained bulk Cu atom. The curvature of $E(h)$ is such that $E(5) + E(7) < 2E(6)$, as shown by the tie line in Fig. 8b. Thus, like the spring model, the direct EAM simulations with no model assumptions also shows that an unclamped film slab has $d^2E/dh^2 < 0$ (Eqn. 1). As the films thicken, $E(h)$ approaches $2\gamma_0$. In words, the slab energy differs from that of bulk metal by the cost of making the two surfaces. The cost of making a film from an infinite bulk becomes smaller as the film gets thinner. In the next section, we show even more directly that this dependence of film energy on thickness results from surface stress.

### 4.5  Elasticity model of de-wetting due to surface stress

The simple spring model given in section 4.3 illustrates the basic reason why surface or interface stress might cause films to be morphologically unstable. However, the model is too simple to be quantitatively compared to the atomistic simulations of sections 4.2 and 4.4. Instead, we reference work by Hamilton and Wolfer that determines the moduli for the bulk monolayer and surface monolayer in Cu(111) films using EAM [50]. Their model is a close analogue of the simple spring model (section 4.3) and is based on work by Gurtin and Murdoch [51, 52]. The model treats the surface layer as a separate layer with different elastic properties from the "bulk" (interior) layers. In our application here, we assume that shear components of the strain are negligible and that the surface-normal components of the stress are zero. Under these conditions, the bulk-monolayer energy of a Cu(111) layer may be written:

$$E_{bulk} = \frac{1}{2}M_{11}(\varepsilon_{11}^2 + \varepsilon_{22}^2) + M_{12}(\varepsilon_{11}\varepsilon_{22}) \qquad (3),$$

where $\varepsilon_{11}$ and $\varepsilon_{22}$ are the bulk strains in two directions parallel to the surface and $M_{11}$ and $M_{12}$ are elastic constants related to the bulk elastic constants of a cubic system ($C_{11}$, $C_{12}$ and $C_{44}$) by a coordinate transformation and by the condition that the surface-normal component of the stress is zero. Here, we use the convention that these quantities are in units of eV/surface atom. Thus, the factor $N$ of section 4.3 is not needed. The energy zero is chosen so that $E_{bulk}$ in the absence of strain is zero.

In a similar manner, the surface-monolayer energy of a Cu(111) surface layer may be written:

$$E_{surf} = \gamma_o + \frac{1}{2}\Gamma_{11}[(\varepsilon_{11} + \varepsilon^\star)^2 + (\varepsilon_{22} + \varepsilon^\star)^2] + \Gamma_{12}(\varepsilon_{11} + \varepsilon^\star)(\varepsilon_{22} + \varepsilon^\star) - (\Gamma_{11} + \Gamma_{12})\varepsilon^{\star 2} \quad (4).$$

Here the $\Gamma_{ij}$ terms are the elastic constants of the surface monolayer. The quantity $\varepsilon^*$ is the residual strain and originates because the in-plane surface bond length differs from the bulk bond length. We use the term surface-monolayer energy to distinguish $E_{surf}$ from the surface energy, $\gamma_o$, which is customarily defined for zero bulk strain. That is, at zero bulk strain, $E_{surf} = \gamma_o$. The surface-monolayer energy, which is defined as $E_{surf}(\varepsilon) = E_{bulk}(\varepsilon) + \gamma(\varepsilon)$, explicitly includes the strain dependence of the surface energy.

For a film with a single thickness, symmetry dictates that the strain will be equibiaxial, i.e., $\varepsilon_{11} = \varepsilon_{22} = \varepsilon$. Then, Eqns. 3 and 4 become:

$$E_{bulk} = (M_{11} + M_{12})\varepsilon_{11}^2 \qquad (5)$$

and

$$E_{surf} = \gamma_o + (\Gamma_{11} + \Gamma_{12})(\varepsilon_{11} + \varepsilon^\star)^2 - (\Gamma_{11} + \Gamma_{12})\varepsilon^{\star 2} \qquad (6).$$

As described in ref. [50], EAM simulations can be used to calculate $E_{surf}$ and $E_{bulk}$, as shown in Fig. 9 for equibiaxial strain. The elastic constants of the surface and bulk layers are then determined by fitting this data to Eqns. 5 and 6, giving: $M_{11}+M_{12}$=19.43 eV/atom, $\Gamma_{11}+\Gamma_{12}$ = 23.69 eV/atom, $\gamma_o$ = 0.4172 eV/atom, and $\varepsilon^*$ = 0.0136. We point out that the minimum in $E_{surf}$ occurs at $\varepsilon$ = -0.0136, showing the tendency of surface atoms to relax to an in-plane spacing 1.36% shorter than that of the bulk atoms.

As in section 4.3, we find the minimum energy configuration by solving $dE/d\varepsilon = 0$. The energy of this configuration is:

$$E(h) = \gamma_0 + \frac{2(h-2)(M_{11} + M_{12})(\Gamma_{11} + \Gamma_{12})\varepsilon^{\star 2}}{(h-2)(M_{11} + M_{12}) + 2(\Gamma_{11} + \Gamma_{12})} - (\Gamma_{11} + \Gamma_{12})\varepsilon^{\star 2} \qquad (7).$$

As in the previous section we solve to get:

$$\frac{d^2E}{dh^2} = -\frac{8(M_{11} + M_{12})^2(\Gamma_{11} + \Gamma_{12})^2\varepsilon^{\star 2}}{[(h-2)(M_{11} + M_{12}) + 2(\Gamma_{11} + \Gamma_{12})]^3} \qquad (8).$$

Since $M_{11}+M_{12}$ and $\Gamma_{11}+\Gamma_{12}$ are positive, we conclude that $d^2E/dh^2$ is negative for all values of $h \geq 2$, showing that the films are unstable with respect to thinning and thickening simultaneously (Eqn. 1).

We can now compare the $E(h)$ determined directly from EAM (filled circles in Fig. 8) with the analytical expression for $E(h)$ (dashed line) using the numerical quantities given above from ref [50] in Eqn. 7. The agreement is good, showing that the thickness dependence of the model-free EAM simulations is accurately described by an analytical model whose essential feature is the strain-dependence of surface stress.

To check whether Cr(110) has elastic properties that are much different than Cu(111), we performed test density functional theory (DFT) calculations with the methods described in reference [34] [i]. The in-plane equilibrium spacings of a 5-layer slab are about 1.8% and 0.9% smaller than the values calculated for bulk Cr in the [110] and [100] directions, respectively. These values are comparable to the strains of Cu(111) slabs of the same thickness, as calculated by EAM [50].

There is even a more stringent test of the elasticity model: we can now directly compare the results of this elastic model (Eqn. 7) to the EAM simulation of the energy change resulting from locally thickening and thinning the slab, as discussed in section 4.2 and shown in Fig. 7. Upon numerical substitution into Eqn. 8 with $h = 6$, we find $d^2E/dh^2 = $ -0.16 meV/atom. Careful examination of our definition for this derivative (Eqn. 1) shows that it corresponds to moving two atoms. So the energy change per atom predicted by the analytical elasticity model is -0.08 meV. This value is in good agreement with the energy change of -0.1 meV/atom found after making 5- and 7-layer film from 6-layer film in the stepped slab (Fig. 7). Thus, our analytical model accurately reproduces the full EAM simulations of a stepped slab in section 4.2. This agreement gives us confidence that surface stress indeed dominates how the energy of unclamped slabs change with thickness. Our elasticity model also shows that including the energies associated with steps is not needed to explain the energies of the stepped slab (Fig. 7). Thus, it appears that step energies, step relaxations, and step-step interactions play minor roles in the energetics of the instability.

---

[i] The Cr calculations performed for this study are non-spin polarized because all experiments are performed well above the Neel temperature (38°C) of Cr.

We calculate that surface stress causes a driving force for mass flow that is on the order of 0.1 meV/atom for Cu(111). This driving force is approximately two orders of magnitude smaller than that needed to nucleate new layers at a measurable rate on ice [18], for example. At first glance, this driving force may seem small. However, the Gibbs-Thomson driving force for Ostwald ripening [53] has similar values for metals. For example, the Gibbs-Thomson chemical potential difference between a metal atom attaching at a straight step edge compared to one detaching from an island of radius 10000 atoms (i.e., a few microns) is also on the order of 0.1meV, assuming a typical close-packed-step energy of 1 eV/atom. Yet ripening or island-stack decay for islands with size of this order of magnitude is observable [54, 55]. Furthermore, as mentioned in section 4.2, the stress of the heterophase interface may be much larger than the stress of the free surface. If so, our estimate based on the latter could underestimate the driving force.

Another complication is that EAM neglects quantum size effects (see, for example, [56]), which could be large compared to these stress-relaxation energies. However, one expects quantum size effects to be oscillatory with film thickness and, thus, do not obviously provide a general mechanism for the instability of arbitrarily thick films (as we observe.)[j]

To summarize section 4, we have shown that surface or interface stress provides a driving force for mass transfer that simultaneously thins and thickens a film. Physically, the film energy depends on thickness because the surface monolayer and the interior layers want to have different in-plane lattice spacings. In very thin films, the energy of the surface monolayer dominates and the interior layers are strained. As the film thickens, the interior layers become less strained but the surface monolayer becomes more strained. The film energy per area increases monotonically but increasingly less rapidly with thickness, as shown in Fig. 8. This thickness dependence, with the second derivative of energy vs. thickness being negative, means that a film can lower energy by simultaneously thinning some regions while thickening others (Fig. 7 and Eqn. 1). This mass flow acting according to the kinetic pathway of Fig. 5 allows the wetting layer (substrate) to be exposed eventually, a critical step in film de-wetting. A fuller treatment of our mechanism would explicitly treat the energy cost of lengthening steps as shown in Fig. 5 (see also sections 3.2 and 6.4) together with the energy reduction coming from thinning and thickening of adjacent film regions (Fig. 8) as well as the mechanisms of mass transport between and along steps. Often such treatments find that step profiles are most unstable to sinusoidal perturbations of a particular wavelength [57]. Thus, such a theory may also account for the well-defined periodicities we often observe emerge during de-wetting, for example, in Fig. 2 (see also section 6.4).

---

[j] A detailed treatment of the Cr(110) system using DFT is problematic because separating elastic energy effects from quantum size effects using the technique is challenging. In addition, the elasticity model of section 4.5 would have to include anisotropy. Finally, a DFT calculation of the *stepped* slab in Fig. 7 would be exacting due to the size of the periodic cell. To avoid these complexities, we have analyzed the model system of Cu(111) slabs simulated using EAM.

## 5. Generality and comparison to other de-wetting mechanisms

### 5.1. Generality of de-wetting mechanism

A phenomenon special to the Cr/W(110) system is that the trenches and 3D islands preferentially form stripes along a particular crystallographic direction of the substrate (see Fig. 2). We emphasize that while this anisotropy makes observing the trench and island formation considerably easier, the de-wetting mechanism in no way requires such anisotropy. In isotropic systems, trenches and fingers will develop along the local direction that allows the film to thicken and thin most rapidly, i.e., along the steepest ascent/descent of substrate steps [10]. The only requirements are that the film/substrate systems be unstable with respect to de-wetting (i.e., the system can lower its energy by simultaneously thinning some regions and thickening others) and that atomic steps are present.

Figures 10 and 11 show the de-wetting in two other film/substrate systems, Ag films on W(110) and Ru(0001). Indeed, in both these systems the thin wetting layers [k] are first exposed at bunches of substrate steps. Once the wetting layer is exposed in the Ag/W(110) system, it grows in size around the step bunches. However, the exposed wetting layer grows only very slowly away from the step bunches in the Ag/Ru(0001) system. The relatively slow growth of the wetting may result from the low interfacial energy of this film/substrate system [39, 40]. Clearly, de-wetting is facilitated by substrate steps in these two systems [l]. We have not been able to characterize the mass flow that occurs as the wetting layer is exposed at substrate steps for Ag on W(110) and Ru(0001) [m]. Given the clear correlation between where the wetting layer is exposed and bunches of substrate steps, we speculate that the mechanism of Fig. 5 also causes the formation of the trenches that expose the wetting layer. The next natural question is about the driving force that causes the wetting layer to be exposed in the initially uniform Ag films of Figs. 10 and 11. Like the Cr/W(110) systems, Ag films on Ru(0001) and W(110) are believed to be essentially strain-free [n]. Our observations of de-wetting in these two additional strain-free systems provide further support for our proposal that the relaxation of strain imposed by the substrate need not be the driving force responsible for exposing the wetting layer.

---

[k] The thickness of the Ag wetting layer is 3 ML on both W(110) [10] and Ru(0001) [58].

[l] With time, the 3D Ag islands will move off of the bunches of substrate steps and onto adjacent, relatively flat regions; for example, the bottom of the pit and around the mesa in Fig. 10. That is, the flat-topped islands stop moving when they are no longer able to become thicker by extending themselves down a staircase of substrate steps, the "down-hill" migration mechanism described in [10, 58].

[m] As mentioned in section 1, the pronounced anisotropy that produces the Cr stripes on W(110) also makes characterizing the mass flow during exposure of the wetting layer considerably easier.

### 5.2. Comparison to other de-wetting mechanisms

A general mechanism for de-wetting has been discussed in the literature, based largely on results of films de-wetting amorphous substrates [14, 60, 61]: first, the substrate is exposed at defects like grain boundaries or impurities. The free edge of the film that bounds the exposed substrate (void) is morphologically unstable because the film has high local curvature at the boundary with the bare substrate [o]. The film edge retracts and becomes thicker. Then, the film edge develops a Rayleigh instability laterally along its length. Locally thinner regions along the edge continue to thin and develop into finger-like voids, which lengthen perpendicular to the original film edge. The film fingers between the voids can then undergo another Rayleigh instability and pinch off a line of film islands.

We see that Cr de-wets W(110) by a different mechanism. Most importantly we observe that the trenches themselves are formed by a "fingering" instability. Once formed, the trenches lengthen both uphill and downhill by removing layers as the instability continues to grow (see section 6.1). We note that the de-wetting mechanism of Fig. 5 is not a type of Rayleigh instability, which lowers surface energy, because total step length increases.

At first glance, our de-wetting mechanism would seem to be irrelevant to crystalline films on amorphous substrates since the latter do not have well-defined steps. Yet some aspects of our focus on atomic steps might be relevant. We note that Burhanudin et al. [62] found that steps on very thin Si(111) films on silica influenced the de-wetting. That is, de-wetting produced Si stripes that could run along three crystallographically equivalent Si directions. However, which direction was preferred was controlled by the orientation of the atomic steps on top of the Si(111) film – the Si stripes formed preferentially in the direction perpendicular to the Si steps. This effect can be interpreted using the concepts described here. If the Si steps advance and retract perpendicular to their edge direction, the film will be simultaneously thickened and thinned in local regions [p]. This motion can provide the concerted mass transfer needed for de-wetting.

---

[n] Ag films on Ru(0001) are known to be strain-relaxed by a dislocation network that does not extend into the second Ag layer [39, 40]. The dislocation network within the first 1-2 layers of Ag on W(110) relaxes the stain so that thicker films are essentially bulk Ag(110) [59].

[o] Note that the operative curvature is not that laterally along the boundary between the bare substrate and the film. Instead the curvature is that of the film abruptly rising from the bare substrate.

[p] Figure 5 can be used to visualize a crystalline film on an amorphous substrate by coloring the red substrate blue, converting it to film. That is, consider that all the material shown in Fig. 5 is crystalline film on top of a flat amorphous substrate. The interface of the film with amorphous substrate is the flat bottom of the stepped slab.

*5.3. Relationship to the de-wetting of films strained by substrates*

In this section, we describe how the mechanism of Fig. 5 might also be relevant to films strained by epitaxy with substrates. As discussed in section 4.1, the ATG theory is commonly used to understand why strained, flat films are unstable relative to undulation. However, the ATG treatment is based on continuum analysis and, thus, ignores the difficulty of creating new atomic surface steps. The simplest extension of the ATG instability to stepped films would follow the spirit of Fig. 1. That is, the film would become undulated by nucleating new steps that form valleys and ridges. But this step nucleation should be energetically costly, as we noted in section 1. However, the mechanism of Fig. 5 also provides a pathway for strained flat films to become undulated. This point is illustrated by comparing the cross sections labeled A'B' and G'H' in Fig. 5. Cross section A'B' is a flat film. After concerted step advancement and retraction, the film is undulated, as in cross section G'H'. Thus, the mechanism of adjacent sections of film steps moving uphill and downhill, respectively, relative to the steps of the substrate provides a kinetic pathway for both strained and unstrained films to become undulated without having to nucleate new film steps. In the ATG case, roughening produces ridges that reduce the strain energy imposed on the film by the substrate. Even when the film lattice is not clamped to the substrate lattice, surface stress strains the unclamped film and provides a driving force for a flat film to become undulated.

## 6. Details of Cr stripe formation

In this section, we provide details about what happens after the cooperative motion of film steps relative to substrate steps has exposed the wetting layer. We describe how the Cr film is converted into an organized pattern of Cr stripes as the trenches that expose the wetting layer lengthen. We also discuss the kinetics of trench motion, the stripe morphology, the mechanism that the Cr stripes use to thicken and narrow, and the factors that control the stripe pattern.

*6.1. Kinetics of trench motion*

After the uphill motion of the Cr steps exposes the Cr wetting layer, the newly formed trench grows rapidly along the W[001] direction. The rate of mass transfer occurring during trench growth is much greater than the rate during the thinning and thickening processes that create the trenches. This difference in mass-transfer rate is dramatic in the videos associated with Figs. 3 and 4. The trench growth is now driven by the interfacial/surface energies of this Stranski-Krastanov system [2, 4]. That is, the thermodynamically stable Cr wetting layer has a low surface energy relative to the surface energy of thick Cr and the energy of the interface between the Cr wetting layer and the thicker Cr layers. Thus, once the wetting layer is exposed, the system can lower its energy by decreasing the area of film/substrate interface, which requires increasing the wetting-layer area and thickening the 3D islands. We next discuss how the trenches grow to make a pattern of Cr stripes.

The trenches expand in both the uphill and downhill directions, as referenced to the staircase of substrate steps (see Fig. 12a). With the exception noted below, the uphill-moving trenches and the downhill-moving trenches move with constant velocity when traveling through uniformly thick film regions. But the downhill velocity is slower than the uphill velocity, as shown in Fig. 12. The velocity difference can be large – over 2.5 times in the example of Fig. 12b. Higher-resolution imaging in Fig. 13 reveals the origin of the slower downhill velocity. The downhill motion is not continuous, unlike the continuous uphill motion. The trenches do not move steadily downhill but instead occasionally stop moving. Analysis reveals that the trenches stop momentarily when they reach descending substrate steps (see Fig. 13). This observation establishes that a barrier exists to further downhill motion once the trench reaches a descending substrate step. As Fig. 14a shows schematically, the barrier arises because a pit must be nucleated next to the descending substrate step. In contrast (see Fig. 14b), a pit does not need to be nucleated for the trench to cross ascending substrate steps because atoms just have to be removed from film step edges. Because of the nucleation barrier, the downhill trench velocity decreases with decreasing separation of substrate steps. While we have not thoroughly examined the effect, we find that the difference between uphill and downhill trench velocity becomes smaller with increasing temperature. Thus, the barrier to nucleating a pit next to a descending step becomes easier to surmount with increasing temperature.

We emphasize that our mechanism of exposing the wetting layer (Fig. 5) avoids nucleating pits in film layers. Once the wetting layer is exposed, the trench can also move uphill without pit nucleation. Downhill motion, however, requires a type of pit nucleation. While the barrier for nucleating next to a descending substrate step is likely to be less than the barrier for pit nucleation away from a substrate step (i.e., on a terrace), the barrier has a measurable effect on trench velocity. That we are able to directly observe the barrier's effect on trench velocity again highlights that discrete (non-continuum) processes are involved in film de-wetting.

Figure 15a shows the dependence of uphill trench velocity on temperature from film regions of the same thickness. (Both uphill and downhill trench velocity decrease with film thickness.) The trench velocity depends strongly on temperature, and, in fact, exhibits Arrhenius behavior with an activation energy of 2.6 ± 0.2 eV (Fig. 15b). For comparison, Allen measured an activation energy of 2.2 eV for the growth of grain-boundary grooves in Cr under conditions established to be controlled by surface diffusion (1400-1700°C) [63]. Given that the grain-boundary grooving experiment was performed on polycrystalline Cr with a preferred (001) orientation, the agreement between the activation energies of trench growth and groove growth is reasonable. Thus, we conclude that the uphill trench growth is limited by the rate of mass transfer by surface diffusion [64], which depends exponentially on the sum of the energies needed to create and migrate the diffusing species, i.e., Cr adatoms. As discussed above, however, the rate of downhill trench growth can be limited by pit nucleation.

## 6.2. Morphology of the Cr stripes

To understand the mechanism that allows the Cr stripes to thicken after they have formed, we must first characterize the stripe morphology. Selected-area low-energy electron diffraction (LEED) shows that the top of the Cr stripes are (110) facets. Analysis also shows that the Cr stripes have well-defined facets on their sides. AFM (Fig. 16) revealed that the angle between the W(110) surface and the sidewall facets of the Cr stripes was about 20°. A more accurate value comes from analyzing how the LEED spots from the sidewalls move with electron energy, as shown in Fig. 17. In a LEEM instrument, diffraction spots from surfaces normal to the incident electron beam do not move on the detector when electron energy is changed [65]. However, spots from surfaces inclined from the beam move with energy. The facet angle is determined by analyzing how the spots move with electron energy [66]. Figure 17 shows the perpendicular momentum transfer, $\Delta k_z = k(1 + \cos\theta_0)$, and the parallel momentum transfer, $\Delta k_x = k \sin\theta_0$, of some facet diffraction spots. Here, $k$ is the incident wave vector and $\theta_0$ is the energy-dependent angle that the diffraction spot makes with the surface normal, as measured after the diffracted electrons are accelerated in the LEEM. $\theta_0$ values were calibrated using the W(110) lattice constant and the separation between the specular and first-order diffraction spots from W along the W[110] direction. The facet angle $\phi$ (see Fig. 18) is determined from the slope $S$ of the lines in Fig. 17 using the geometric relationship described by Fig. 3 of ref. [66] -- $\phi = \pi/2 - \tan^{-1}(S)$. We find that the stripe sidewalls make an angle $\phi = 18.6 \pm 0.7°$ with the substrate. This value establishes that the sidewall facets are Cr {120} planes, which make an angle of 18.4° with the (110) plane. Fig. 18 gives a schematic illustration of the stripe geometry. Stripes formed during the homoepitaxial growth of W(110) and Mo(110) at relatively low temperature also have {120} facets on their sidewalls [29].

## 6.3. How developed stripes thicken

Once formed, the Cr stripes become narrower with annealing, as shown in Fig. 19. At the same time, the stripes become thicker, using the mechanism illustrated in Fig. 20. The thickening occurs by film steps moving along the stripe tops, in the direction that is down the staircase of substrate steps (Fig. 20a). The net process transfers Cr from the sides of stripes to their tops. The stripe shape of the islands, which results from the anisotropy of the Cr/W(110) system (see section 7.2), allows thickening to occur very efficiently. There are many film steps along a stripe and the Cr stripe is locally thickened each time a Cr film steps advances over a descending substrate step. The Cr stripes can easily thicken (and narrow), without having to nucleate new steps, as long as Cr steps exist that can advance downhill (see Fig. 20).

The stripe morphology is remarkably stable. Only after long times at high temperatures do the stripes begin to breakup. Failures occur first at the "y" junctions where trenches terminate inside a stripe (Fig. 19 shows examples). Only when the temperature is high enough for sublimation do non-bifurcated stripes themselves segment into more compact but still elongated shapes. These more-compact islands continue to evolve, thickening by the "down-hill" migration mechanism of island motion that we have previously described [10, 58]. Indeed, compact 3D islands only thicken by moving their entire mass down the staircase of substrate steps, a relatively slow process. Still, the underlying thickening mechanism is the same for both compact and stripe-shaped islands -- film steps advance downhill. Because the islands maintain nearly flat tops, they thicken as their film steps advance relative to the fixed substrate steps. Our observations that the Cr stripes continuously evolve establish that they result from kinetic processes and are not equilibrium structures. (The equilibrium island shape is likely closer to the compact islands observed at high temperature.)

*6.4. Pattern formation*

In this section, we discuss some of the factors controlling the straightness, separation, and height of the stripe patterns formed when Cr films de-wet W(110). Figure 21 shows the effect of initial film thickness. The stripes are very well aligned along the W[001] direction for films roughly thicker than 20 ML. Thinner films still form stripes but the stripes get shorter as the films get thinner and the stripes are less well-aligned along the W[001] direction. In the thinnest films, some of the stripes lay along W steps that run close to the W[001] direction. For thinner films, trenches exposing the wetting layer are created with high density on the substrate. Whereas, with increasing film thickness these trenches form only rarely, at locations of very high substrate step density.

These observations are fully consistent with the de-wetting mechanism proposed in Fig. 5. To expose the wetting layer, film steps must cooperatively advance and retract over the same number of substrate steps as the film thickness in atomic layers minus the wetting layer. For thin films, many locations on the substrate have this number of closely spaced substrate steps. For thicker films, however, to expose the wetting layer, film steps must retract and advance over more substrate steps. The substrate regions that have this number of closely spaced substrate steps become rarer with increasing film thickness.

Given sufficient annealing time, the trenches advance, exposing more wetting layer, until they meet substrate already de-wetted by another trench. The length of a given stripe, then, is controlled by the separation between the initial nucleation events. The stripe length can be quite long in thick films. In the image montage of Fig. 22, for example, trenches nucleated at the substrate step bunches in the upper left and lower right images. Continuous stripes that are about 360 μm long span the two nucleation sites. Thus, the de-wetting mechanism can produce a striped pattern over remarkable large distances.

Figure 23 shows that stripes with a fairly regular separation form at a uniform step bunch. The separation (period) of the Cr stripes will in general depend upon the film thickness, the density of the substrate steps at the nucleation site and perhaps upon the orientation of the substrate steps relative to the stripe direction. For a given film thickness, the stripe period is determined by the energy balance between the cost of lengthening film steps (section 3.2) and the energy reduction resulting from simultaneously thinning some film regions while thickening other regions (see sections 4.4 and 4.5 and Fig. 8). Fingering instabilities such as observed in Fig. 23 have been found in a variety of de-wetting film/substrate systems, including combinations of metals on crystalline substrates [15], metals on amorphous substrates [14], and semiconductors on amorphous substrates [11, 61, 62, 67].

After long annealing the Cr stripe separation is not only determined by the instability at the local nucleation source but also by stripes that form at the closest trench-nucleation site along the [001] direction. This effect is clearly seen in Fig. 22 where stripes are subdivided by trenches moving in opposite directions. Some trenches terminate on their right, showing that they were moving to the right. Similarly, some trenches terminate on their left, showing that they were moving to the left. Figure 12a also shows stripe interdigitation, which can reduce the initial separation by a factor of two. The stripe height is initially controlled by the film thickness. The initial stripe height is close to the original film thickness. But because the stripes can continuously thicken (see Fig. 19), their height and width are ultimately controlled by the annealing time (see Fig. 16).

Finally, we note that Cr will form stripes along the W[001] direction during de-wetting an initially uniform film of sufficient thickness even when the substrate steps lie close to the [001] direction. Figure 24 shows an STM image from a different W(110) crystal than the other figures. The Cr stripes are aligned along the [001] direction even though the substrate steps, imaged in the conformal wetting layer, also lie nearly along [001]. We speculate the trenches that expose the wetting layer formed at some minority of step bunches that were inclined somewhat from the [001] direction, using the pathway shown in Fig. 5. Once a trench nucleates, it can lengthen by transferring Cr in the direction perpendicular to the W steps that points down the staircase of substrate steps. The practical conclusion is that Cr stripes produced by de-wetting sufficiently thick films are aligned along the [001] crystallographic direction, independent of the substrate step direction.

Summarizing pattern formation, unidirectional stripes can be made in the Cr/W(110) and Fe/W(110) systems relatively easily [20-22]. The stripe separation can be controlled somewhat with film thickness even on natural (non-patterned) substrates. More control could be achieved by engineering spatial variations in the substrate step density using, for example, lithography or ion milling [10].

## 7. De-wetting during film deposition and the origin of anisotropy

### 7.1. De-wetting during film deposition

In addition to characterizing how initially uniform films, deposited at low temperature, de-wet upon annealing, we have also studied how de-wetting occurs *during* film deposition at elevated temperatures. Figure 25 shows that the first three Cr layers grow in perfect step flow [33, 34]. But as soon a 3rd layer step reaches a descending substrate step, the higher layers grow very rapidly. (See the video associated with Fig. 25.) That is, after a 3rd layer has reached a descending substrate step, 4th layer Cr forms on the lower terrace and rapidly advances. After this 4th layer has crossed the terrace and reached the next descending substrate step, 4th and 5th layers rapidly start to grow on the next lower terrace. This growth mechanism, which also occurs in isotropic systems, has been explained in the literature [10, 58]: stress-relieving dislocations are introduced into the Cr film when the 4th layer forms (see section 7.2). Thus 4-layer films have lower energy than 3-layer films. Therefore, 4th-layer regions provide a preferred sink for incoming atoms. When 4th-layer steps advance over the next descending substrate step, the local thickness increases to 5 layers. Thus, the substrate steps enable a nucleation-free process during film growth that leads to local film thickening and formation of mesas (wedges) [q].

We are now able to compare the energies of the first few Cr layers with the energies of thicker Cr layers. Figure 25 shows that 4th-layer Cr grows by consuming 3rd-layer Cr. While not shown in Fig. 25, over longer times, 2nd-layer Cr is also consumed during annealing. Thus, the 2nd and 3rd layers are less stable than the 4th layer. As discussed in section 4, the 4th layer is essentially decoupled from the substrate. Thus, surface stress causes thicker layers to have increasingly higher energy (Fig. 8). Combining this information leads to the schematic energy dependence shown in Fig. 26. Film growth must proceed through the 2nd and 3rd layers, even though these layers are unstable compared to other thicknesses.

Mesas similar to those in Fig. 25 have been observed by Shvets and co-workers for Fe and Cr growth on Mo(110) [31, 68]. However, they propose that "wedge" thickening occurs from new layers nucleating on existing film layers. They also propose that upward mass flux needed for new-layer nucleation is supplied by enhanced diffusion along the cores of edge dislocations at the island edges [31]. Our observations of 3D-island formation during film growth are consistent with the mechanism of Ling et al. [10, 58]. Specifically, we do not observe that 3D islands thicken by new layers nucleating on top of existing layers (Fig. 1) [r]. Once any area thickens to more than three layers by step-flow growth, these regions thicken into 3D Cr islands as their film steps advance down the staircase of substrate steps (see schematic in Fig. 25). We note that including the important role of substrate steps in models of stripe formation during growth is a challenge [69].

---

[q] The islands that form during elevated-temperature growth are quite elongated along the [001] direction when the substrate steps are perpendicular to this direction, as in Fig. 25. When the substrate steps are inclined to the [001] direction, the 3D Cr islands tend to follow the step direction, similar to the de-wetting of thin films (see Fig. 21).

### 7.2. Origin of anisotropy in Cr/W(110) system

Again, the only feature that is obviously special to the Cr/W(110) system is the anisotropy that elongates the 3D islands appreciably along the W[001] direction. The observations of de-wetting during growth give insight into the origins of the anisotropy. The anisotropic islands produced during high-temperature growth do not become more compact upon annealing at the growth temperature. This observation suggests that the kinetic effect of anisotropic diffusion during growth is not the cause of the elongated island shape [26, 31]. Instead, we suggest an anisotropic step or strain energy is responsible. The growth becomes anisotropic as soon as the 4th layer forms. That is, 4th layer preferentially makes steps along the W[001] direction (see Fig. 25). One possibility is that the Cr steps have lowest energy along the W[001] direction. Another possibility is that the dislocation network that forms in the Cr film when it is 4 layers thick relaxes strain better along the W[001] direction than along the perpendicular direction. Figure 27 shows selected-area LEED from 4 ML Cr (Fig. 27b) and 10+ML Cr (Fig. 27d). The presence of superstructure diffraction spots from both these film thicknesses show that they are non-pseudomorphic and, therefore, considerably less strained than layers 1-3 [s]. But, the superstructure diffraction spots of 4 ML Cr are more intense along the [001] direction than perpendicular to it (see Fig. 27b). This suggests that strain relaxation is not isotropic in 4 layer films. In contrast, the superstructure spots of 10+ ML Cr have uniform intensity along and perpendicular to [001] (see Fig. 27d and [119]), which suggests isotropic stress relaxation. We end by noting that while the {120} planes that exist on the sidewalls of the Cr stripes might stabilize the stripes, the anisotropy in the system manifests itself during de-wetting during growth (Fig. 25) long before sidewall facets exist.

---

[r] However, new-layer nucleation might occur at higher deposition fluxes.
[s] Superstructure spots are observed in films between 4 to about 10 ML thick, as discussed in footnote d.

## 8. Summary

We have analyzed how a flat Cr film devoid of threading, screw dislocations de-wets a W(110) substrate. We have observed in real-time the mechanism that exposes the stable wetting layer, the critical process needed to initiate the conversion of a uniform film into 3D islands during annealing. The mechanism uses the concerted motion of adjacent film steps, which move both uphill and downhill relative to the staircase of atomic steps on the substrate (Figs. 3-5). Some film regions thicken. Adjacent regions thin and eventually expose the wetting layer.

The Cr film on top of the dislocated Cr near the substrate (at most layers 1-3) is essentially unstrained by the substrate. In contrast, most models describing morphological instability consider films that are strained by lattice matching with the substrate. Section 4 examined the energetics of films not strained by substrates, revealing that surface and interface stress can provide a driving force for a flat film to be unstable relative to being undulated. Both atomistic simulations and analytic models show that surface stress yields the dependence of film energy with thickness (a negative second derivative, Fig. 8)) required for the observed simultaneous film thickening and thinning. Thus, film (interface) stress provides a tentative explanation of the instability.

The kinetic pathway of film steps moving relative to the substrate steps can enable de-wetting in any crystalline film/substrate system that contains atomic steps and for which de-wetting is energetically favorable, as for Cr/W(110) (Fig. 2), Ag/W(110) (Fig. 10) and Ag/Ru(0001) (Fig. 11). The mechanism can provide a kinetic pathway for both clamped and unclamped films to become undulated without having to create new steps. The mechanism shown in Fig. 5 also provides a kinetic pathway for the Asaro-Tiller-Grinfeld instability (i.e., a clamped film strained by the substrate). In addition, the mechanism may also have relevance even to the de-wetting of crystalline films on amorphous substrates (section 5.2).

In addition, we have studied in detail the kinetics and pattern formation that occur as the exposed wetting layer grows (section 6). Relatively uniform arrays of Cr stripes are directly achieved during annealing, with the stripe length determined by the separation between nucleation sites (i.e., bunches of substrate steps, Fig. 22). However, the stripes continue to evolve, thickening and narrowing with annealing (Fig. 19) by film steps advancing down the staircase of substrate steps (Fig. 20). Finally, we have studied the de-wetting and 3D island formation that occurs directly during film growth at elevated temperature (section 7). The mechanism of film steps crossing descending substrate steps produces the mesas (wedges) observed after the growth of many non-wetting film/substrate systems [58]. (See, for example, references [70-75].)

Our findings provide insight into the stability of film/substrate systems. For example, the time needed to de-wet a uniform film will increase dramatically with decreasing density of substrate steps. Thus, the practical stability of flat films produced by, for example, low-temperature deposition followed by annealing, can be greatly enhanced by reducing the density of substrate step bunches (Fig. 22). Similarly, 3D island formation during growth at elevated temperature can be avoided by reducing the substrate step density. Recognition of this effect has allowed the synthesis of uniformly flat films of thickness not previously thought possible by step-flow growth [76].

**Acknowledgement** We thank W. G. Wolfer for helpful discussions. This work was supported by the Director, Office of Science, Office of Basic Energy Sciences, Materials Sciences and Engineering Division, of the U.S. Department of Energy under Contracts No. DE-AC04-94AL8500 and DE-AC02-05CH11231, through funding from the Comunidad Autónoma de Madrid and the CSIC through Project No. CCG07-CSIC/MAT-2030, and by the Spanish Ministry of Science and Innovation through Project No. MAT2006-13149-C02-02.

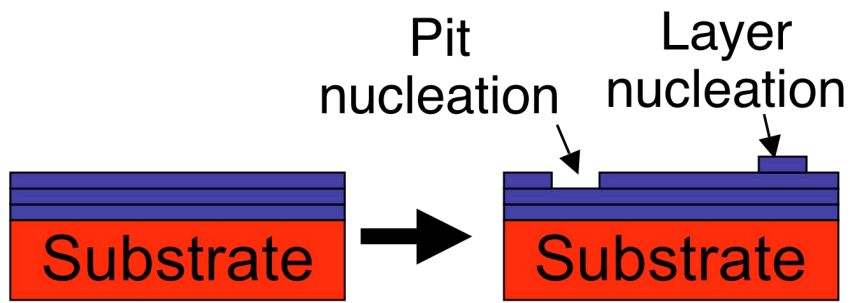

Pit nucleation

Layer nucleation

Fig. 1. Schematic illustration of thickening and thinning, respectively, a single-crystal film by nucleating a new atomic layer or an atomic pit.

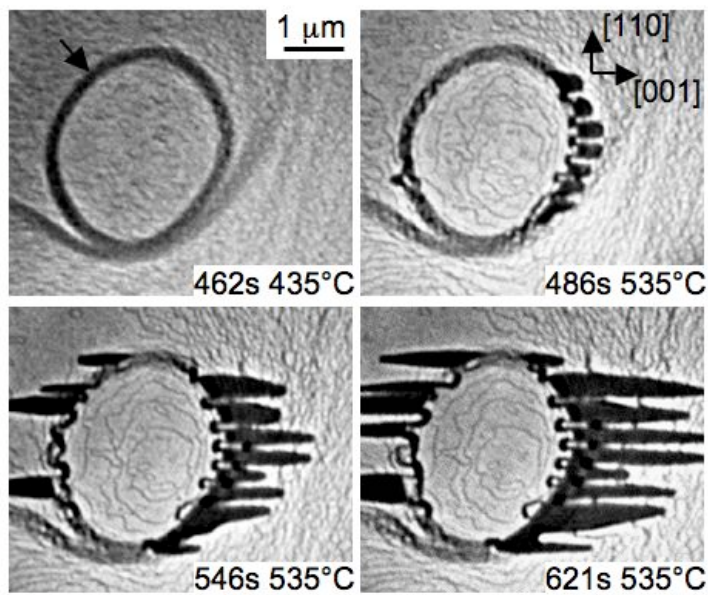

Fig. 2. A time sequence of LEEM images showing how trenches that expose the Cr wetting layer form at step bunches on the W(110) substrate. The circular feature is a pit in the substrate. The dark band surrounding the pit is a bunch of atomic steps in the substrate, which is marked by an arrow at one location. During annealing, the 22 ML Cr film becomes nearly conformal with the substrate. Then trenches that expose the 1 ML wetting layer, the dark horizontal stripes, form at the step bunch that bound the pit.

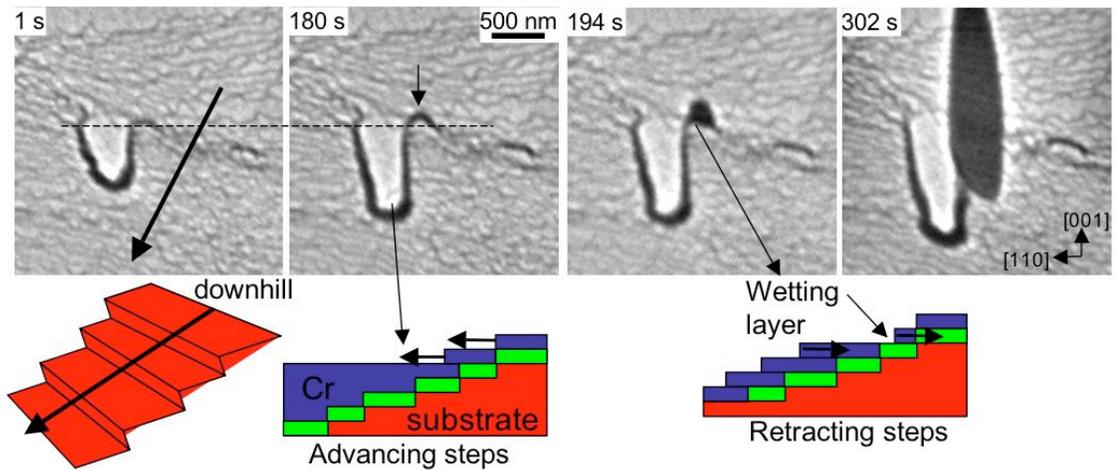

Fig. 3. LEEM images showing how an instability exposes the wetting layer (green). During annealing at 523°C, the flat-topped Cr "finger" left of center grows as Cr steps extends themselves down the staircase of substrate steps (see schematic). The arrow in the first image shows the substrate's downhill direction. A trench, the dark vertical band, nucleates on the finger's right side, where the Cr film is thinner. The horizontal line, fixed at constant substrate position, shows that the Cr steps on the right side of the finger move uphill prior to trench nucleation. This uphill motion locally thins the film and a cusp-shaped bunch of Cr steps forms (see arrow in the 180 s image). At 194 s, the retracting step bunch exposes some wetting layer, which rapidly expands. Cr removed from the trench adds to the finger. A video version is available, Fig3.avi.

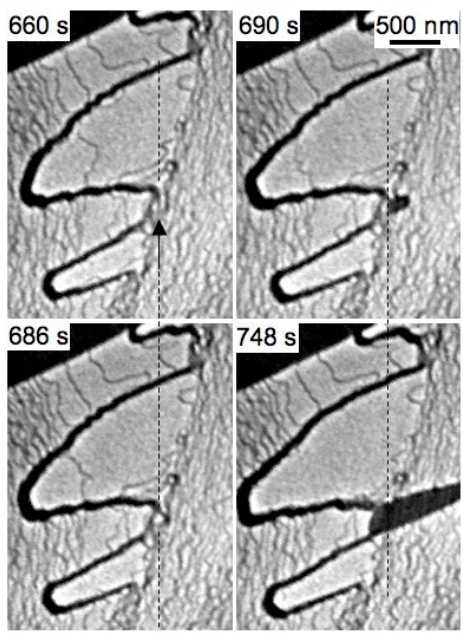

Fig. 4. LEEM images showing an additional example of trench nucleation. During annealing the 25-ML film at 556-568°C, the flat-topped Cr "fingers" grow and thicken as Cr steps extends themselves down the staircase of substrate steps. Simultaneously, the region between the fingers is thinning -- the vertical dashed lines mark the original location of a bunch of Cr steps, the dark vertical band marked by the arrow in the 660-s image. This step bunch retracts uphill (686 s) and exposes the wetting layer (dark patch in 690-s image), which rapidly expands uphill. The uphill movement (retraction) of the Cr step bunches is readily seen in the video version, Fig4a.avi. A final example in video form, Fig4b.avi (594°C, 1.6 μm × 2.6 μm), shows the uphill retraction of monatomic Cr steps as well as step bunches.

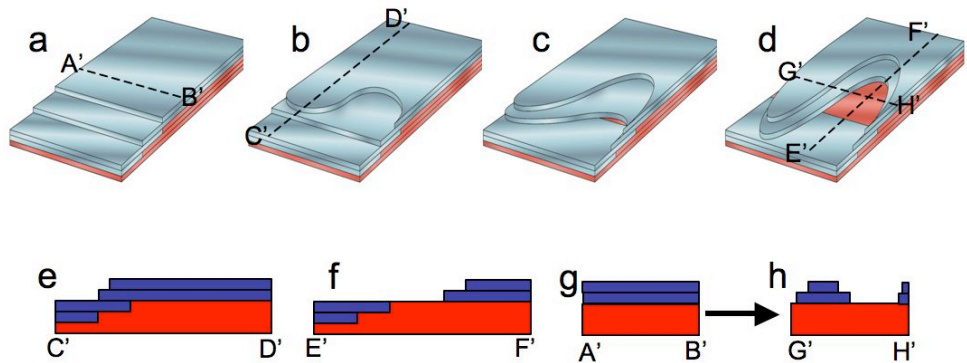

Fig. 5. Schematic illustration of the kinetic pathway that enables a film to de-wet. a). The film (blue) is conformal on the substrate plus wetting layer (red). b). A film step advances over a descending substrate step. This process thickens the region. Concurrently, the film step retracts in adjacent regions. In c and d the instability grows as steps in adjacent regions continue to advance and retract, respectively. Eventually, the retracting steps expose the substrate (or wetting layer). The exposed wetting layer grows rapidly in area to reduce the area of the film's interface with the substrate (wetting layer). Panels e-h show cross sections through the cuts labelled in the 3D models. e shows how film step motion causes local thickening to 3 layers, while f shows local thinning that exposes the substrate (wetting layer). The flat film in g is converted to the undulating film in h.

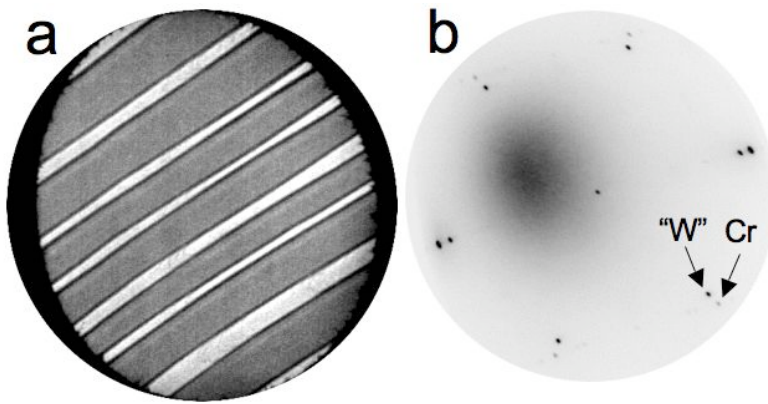

Fig. 6. a). LEEM image showing Cr stripes (bright) interdigitated with wetting layer (dark). A mechanical aperture limits the extent of the illuminating electron beam on the surface. Field-of-view is 7 μm. b). LEED pattern at 47 eV from the region of part (a) showing two sets of diffraction spots, one from the W substrate plus wetting layer ("W"), the other from the thick Cr.

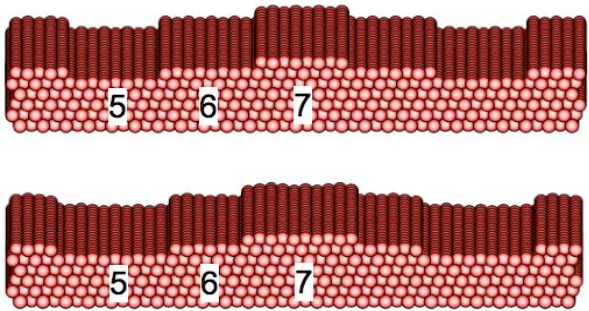

Fig. 7. Two configurations for a stepped Cu(111) film used to evaluate how film energy changes with thickness. The top configuration has terraces of 5-, 6-, and 7-layer-thick Cu. These terraces all have the same area. The bottom configuration is generated by moving two rows of atoms from the edge of the 6-layer terraces to the edges of the 7-layer terrace. Since the number of atoms in both configurations is equal, the energies calculated using EAM can be compared directly. The bottom configuration has lower energy.

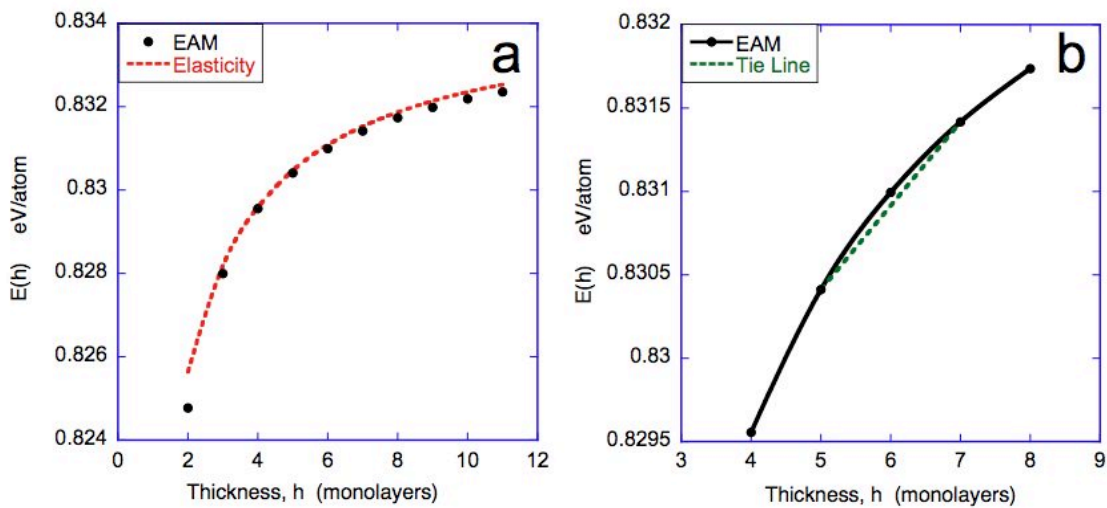

Fig. 8. Film energy, $E(h)$ (in eV/surface atom), for relaxed films as a function of film thickness. a) Films with 2 to 11 layers. Filled circles are from EAM and the dashed line is from the elasticity model of section 4.5. b) Films with 4 to 8 layers, with tie-line construction illustrating how the total energy of a 6-layer film can be reduced by converting to regions of 5 and 7 layers (Fig. 7). The energy is referenced so the energy of a bulk atom in unstrained Cu is zero. As $h \rightarrow \infty$, $E(h) \rightarrow 2\gamma_0$ expressed in eV per atom of surface area. The factor 2 corresponds to the fact that the slab has a bottom and top surface.

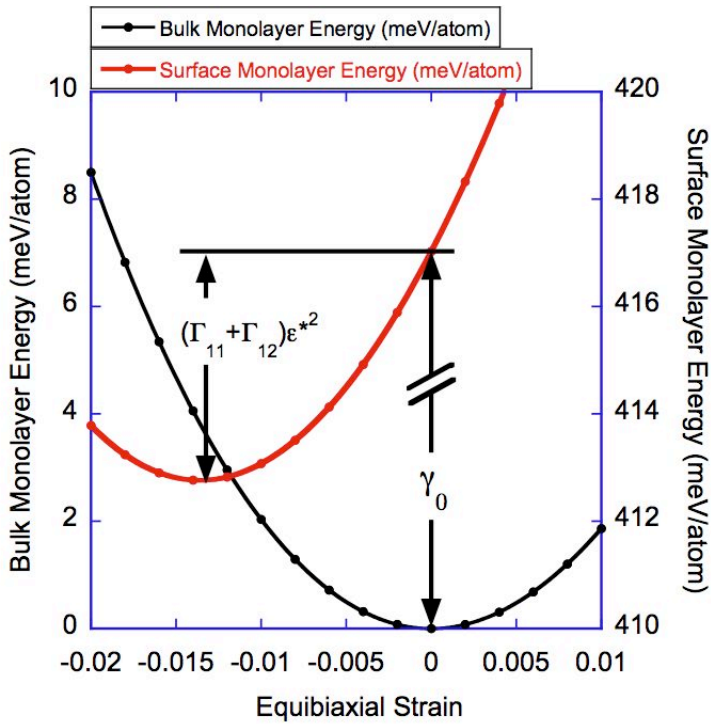

Fig. 9. $E_{bulk}$ and $E_{surf}$ calculated using EAM for equibiaxial strain of Cu(111). These two functions are linked by the definition $E_{surf}(\varepsilon)=E_{bulk}(\varepsilon) + \gamma(\varepsilon)$. Note: because $E_{bulk}$ and $E_{surf}$ are plotted on different y axes, the quantity $\gamma_o = 0.417$ eV/atom is not drawn to scale. The quantity $(\Gamma_{11}+ \Gamma_{12})e^{*2}$ is the energy required to stretch the surface monolayer from its preferred in-plane lattice constant with $\varepsilon_{xx}= \varepsilon_{yy}= -0.0136$ to the preferred lattice constant of the bulk solid, $\varepsilon_{xx}= \varepsilon_{yy} = 0.0$.

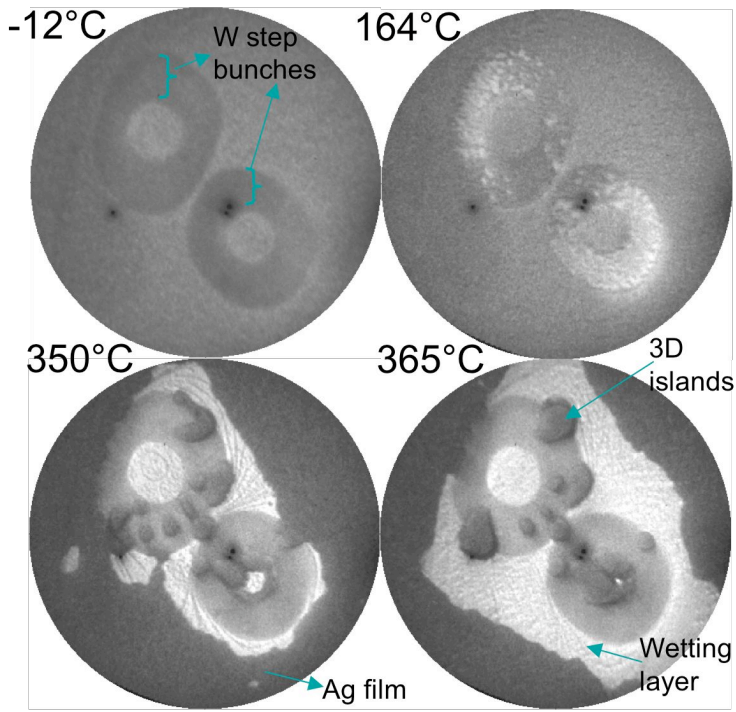

Fig. 10. A 24 ML Ag film de-wetting W(110). The 3-ML wetting layer is exposed at the two large, circular bunches of substrate steps. 3D Ag islands form and grow along the sidewalls of the substrate mesa (upper feature). 3D islands migrate down the steps that bound the pit (lower feature), nearly covering its flat bottom. Field-of-view is 6.7 μm.

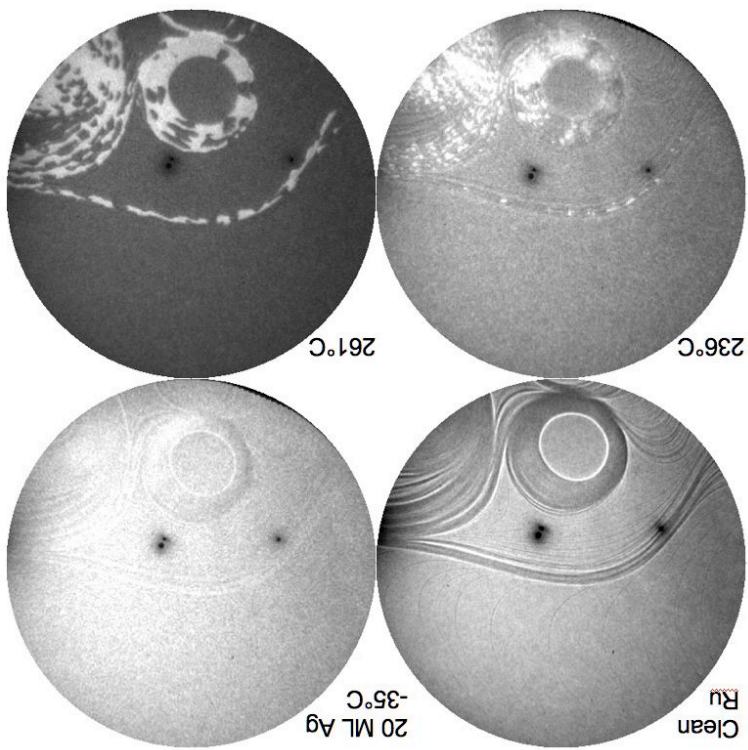

Fig. 11. A 20 ML Ag film de-wetting Ru(0001). Before Ag deposition (upper right image), Ru step bunches image as dark bands. With annealing, the 3-ML wetting layer (bright patches) is exposed at these step bunches. Field-of-view is 19 μm.

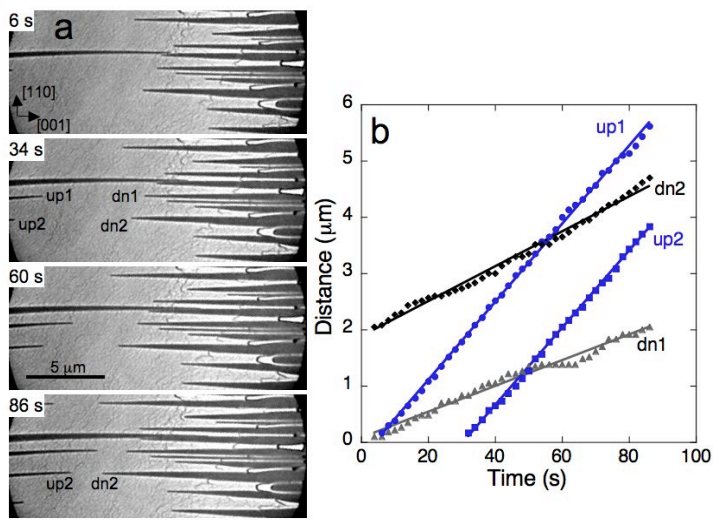

Fig. 12. The trenches that expose the wetting layer move faster up the staircase of substrate steps than down. a). Time sequence of LEEM images showing the growth of trenches (dark, horizontal bands) at 593°C. The trenches moving right on average are going up the staircase of substrate steps. b). Position of the trench ends moving uphill (up1 and up2) or downhill (dn1 and dn2). The trenches move uphill roughly 2.5 times faster than downhill. A video version is available, Fig12.avi.

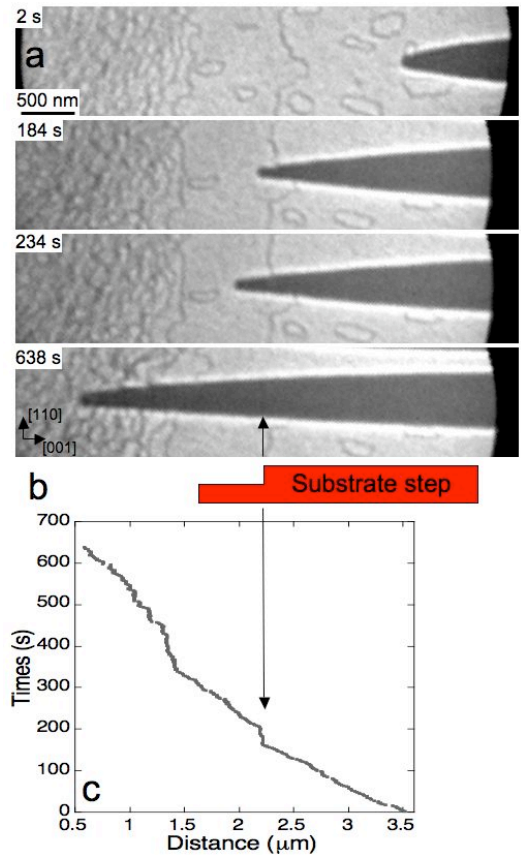

Fig. 13. Non-constant velocity of a trench moving down the staircase of substrate steps. a). Time sequence of LEEM images tracking a trench (dark, horizontal band) that exposes the Cr wetting layer moving downhill at 535°C. In the 638-s image, the arrow marks the location of a step in the substrate and in the conformal 1 ML wetting layer. b). Schematic illustration of the substrate in cross section showing the topography along the trench -- the substrate and conformal wetting layer have an isolated step at the marked position. c). Position of the moving trench end versus time. When the trench reaches the substrate step at about 160 s, it stops moving for over 40 s. A video version is available, Fig13.avi.

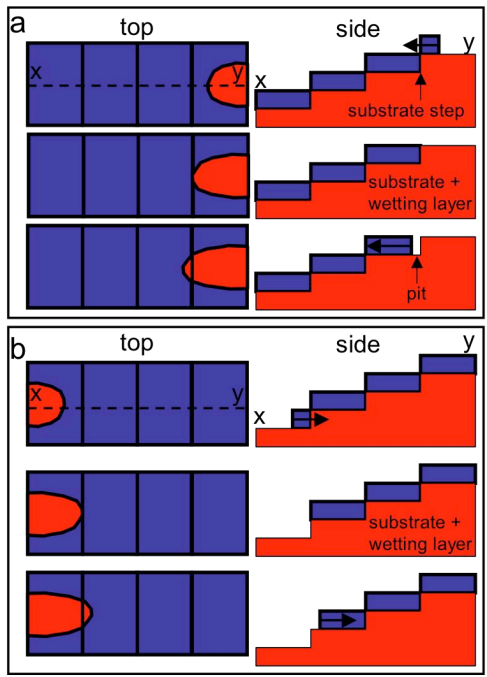

Fig. 14. Schematic illustration explaining why the trenches that expose the wetting layer grow more slowly down the staircase of substrate steps than up the staircase. A one-layer film (blue) on a wetting layer (red) is illustrated. a). Trench moving *down* the staircase of substrate steps. When the trench reaches a descending substrate step, a pit in the lower film layer must be nucleated. b). Trench moving *up* the staircase of substrate steps. When a film step reaches an ascending substrate step, the film layer on the adjacent, higher terrace is simply etched from its edge. Because uphill motion does not require nucleating a pit, the uphill velocity is higher than the downhill velocity.

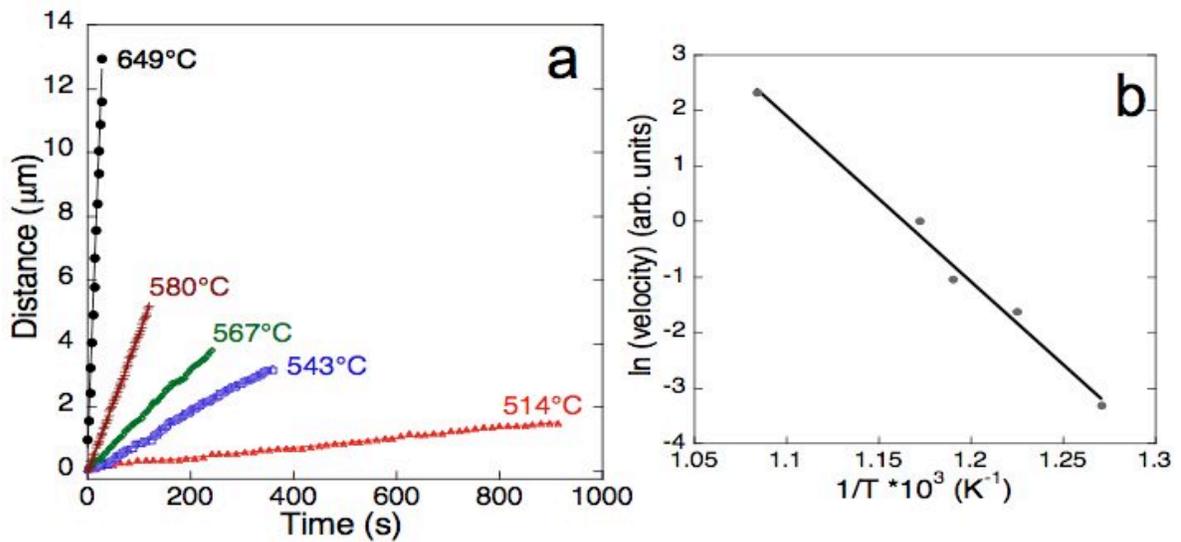

Fig. 15. Temperature dependence of velocity that trenches move uphill in a 21 ML Cr film. a). Position of trench ends versus time for five temperatures. b). Arrhenius plot of the trench velocity, which gives an activation energy of 2.6±0.2 eV.

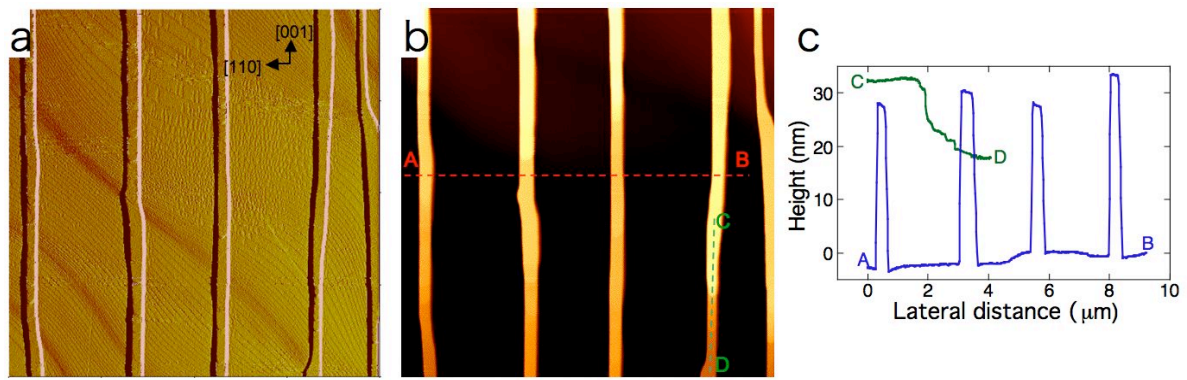

Fig. 16. Ambient atomic force microscopy (AFM) characterization of Cr stripes produced by annealing a 30 ML film for 620°C for about 45 minutes. a). Amplitude of the tapping-mode cantilever oscillation. The Cr stripes are vertical and the faint, curved, diagonal lines are the atomic steps in the 1ML Cr wetting layer. The substrate (and the conformal wetting layer) steps downhill from upper right to lower left. b). Topography image. c). Height profiles perpendicular (A-B) and along (C-D) the Cr stripes. The stripes have flat tops about 250 nm in width. Their heights, about 30 nm, are about 5 times thicker than the starting 6.1-nm-thick film, consistent with the roughly 1/5 coverage after de-wetting. Along their lengths, the stripes have flat-topped segments separated by step bunches and monatomic Cr steps.

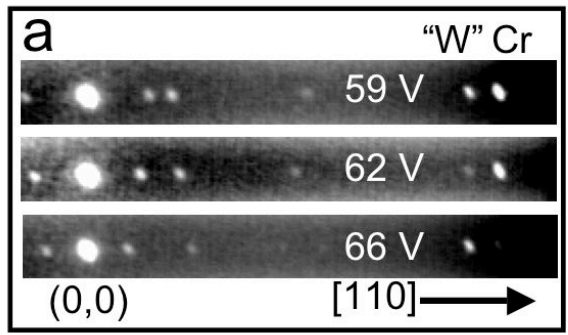

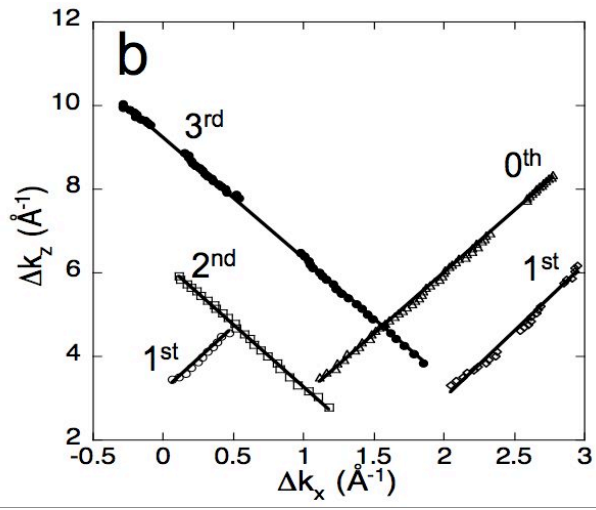

Fig. 17. a). Sections of LEED patterns at three electron energies obtained from the region of Cr stripes shown in Fig. 6a. The (0,0) (specular) beam from the W(110) substrate and pseudomorphic Cr wetting layer is labelled along with their first-order diffraction spot (marked "W") along the [110] direction. Also labelled (marked Cr) is the first-order diffraction spot from the strain-relaxed (110) top facet of the Cr stripes. Unlike the top-facet spots, the weaker spots from the sidewall facets move with energy. b). Perpendicular momentum as function of parallel momentum for sidewall-facet spots of Cr stripes. The lines correspond to zero-, first-, second-, and third-order diffraction, as determined from the y intercept. Breaks correspond to where sidewall spots move through stationary diffraction spots.

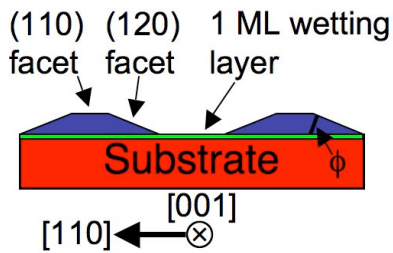

Fig. 18. Schematic illustration of the cross section of the Cr stripes (not to scale) and sidewall facet angle $\phi$.

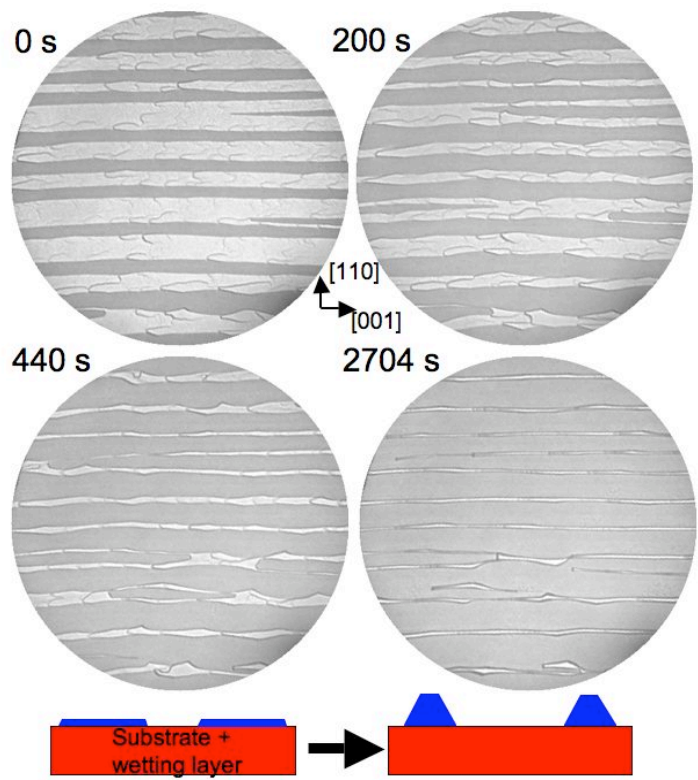

Fig. 19. Time sequence of LEEM images showing the evolution of Cr stripes at 620°C in a 22 ML film. The Cr stripes (bright) become narrower and higher, as shown in the schematic cross-sections. Field-of-view is 19 μm. A video version is available, Fig19.avi.

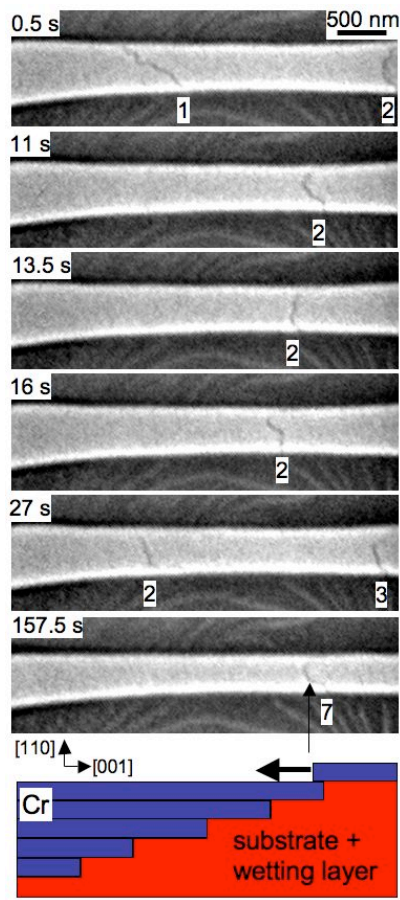

Fig. 20. Time sequence of LEEM images showing the thickening of a Cr stripe (bright) at 630°C. Monatomic Cr steps, the numbered dark lines that run roughly vertical, move from left to right on top of the Cr stripe. Because this direction is on average down the staircase of substrate steps, the advancing Cr steps thicken the stripe (see schematic). The stripe narrows as it thickens, as readily seen by comparing the 0.5- and 157.5-s images. A video version is available, Fig20.avi.

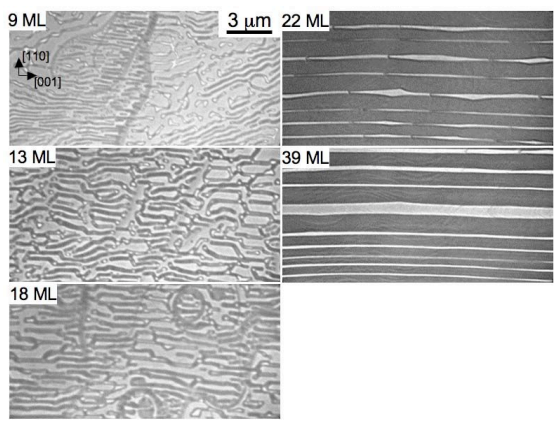

Fig. 21. Dependence of de-wetting behavior on film thickness. The trenches that expose the wetting layer form at high density for thin films and at very low density for thick films. For films roughly thinner than about 20 ML, trenches formed relatively uniformly over the surface because the average density of substrate steps is large enough for the de-wetting instability (see Fig. 5) to expose the wetting layer. For the thickest films examined (about 40 ML), trenches only formed at the rare locations of extremely steep step bunches. Even this low density of trenches was able to de-wet the entire surface given sufficient time – eventually trenches that were created far away would move into a region devoid of the steep step bunches needed to nucleate the trenches.

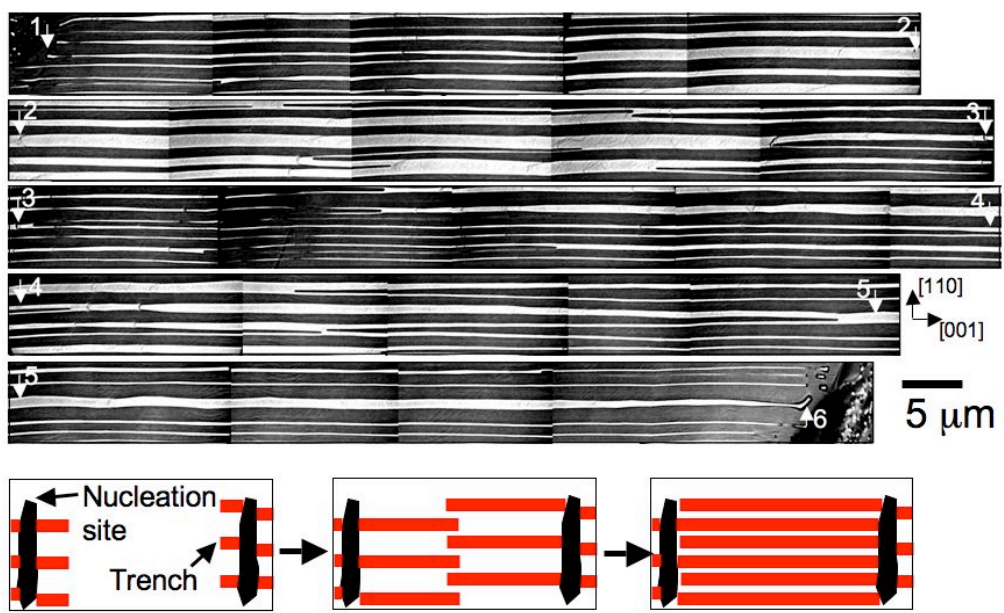

Fig. 22. Example of the Cr length resulting from de-wetting a thick (39 ML) Cr film. Composite of LEEM images tracking a stripe-shaped Cr island that is 360 μm long. Equivalent points in successive rows are marked by the numbered arrows. Within this region, trench nucleation only occurred at the steep bunches of substrate steps near points 1 and 6. Trenches moved both to the left and the right during annealing. The schematic shows the interdigitation of trenches nucleated at separated sites.

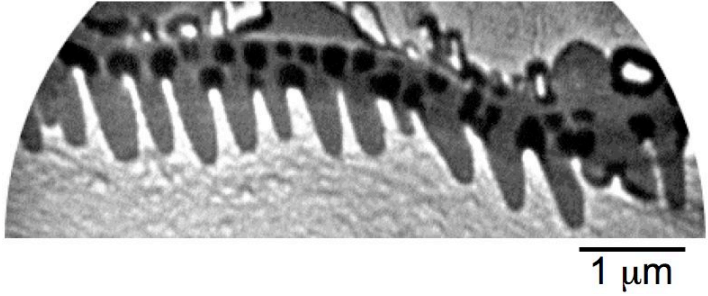

1 µm

Fig. 23. Example showing that a relatively uniform spacing of Cr stripes (bright) forms at a uniform bunch of substrate steps.

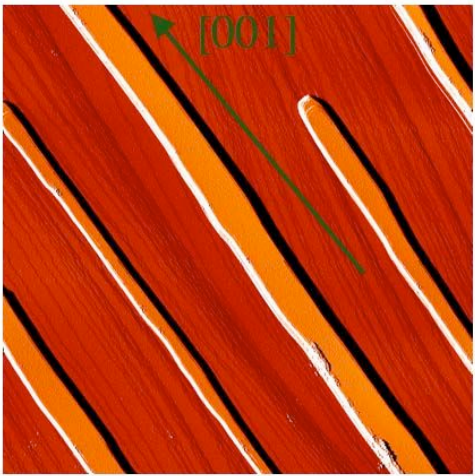

[001]

Fig. 24. STM image (3 µm field-of-view) of an ~10 ML Cr film annealed at 850°C for 10 minutes. Note that the flat-topped Cr stripes are well-aligned along the W[001] direction even though the substrate steps, visible around the stripes, lie close to this direction.

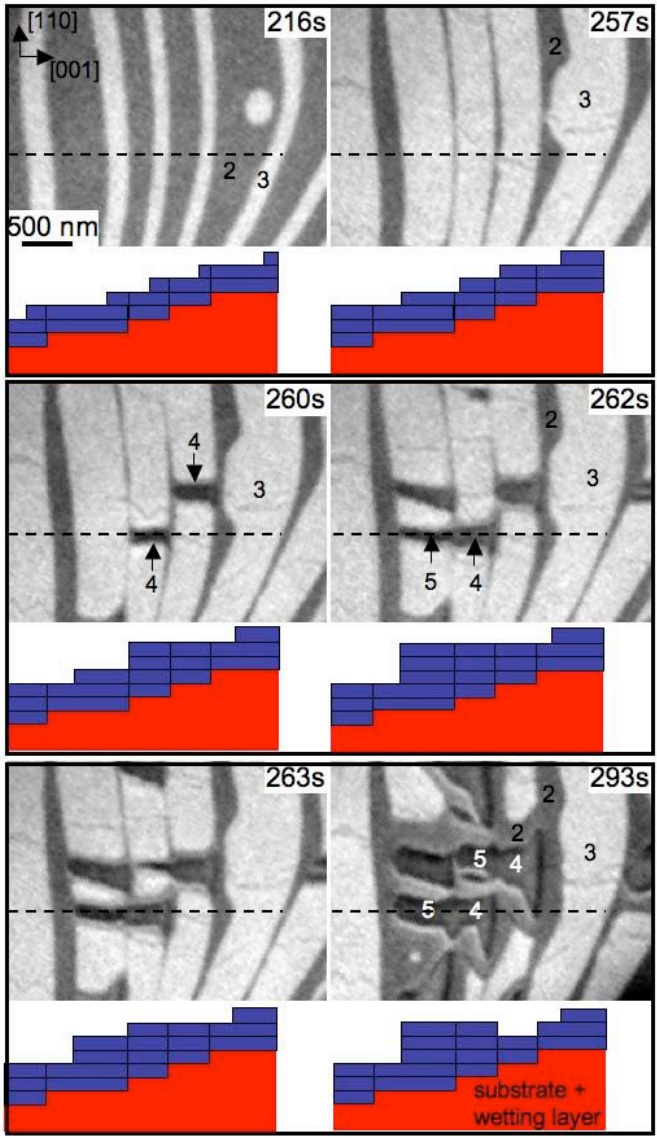

Fig. 25. 3D island formation during film deposition at 360°C. The layer thicknesses are numbered. The substrate steps downhill from right to left. Layers 1-3 grow by nearly perfect step-flow. The 4th layer nucleates when advancing 3rd layer Cr steps reach descending substrate steps. The 4th layer grows rapidly after nucleating, consuming 3rd layer Cr. The growth flux was turned off at 261s. During annealing, the 2nd layer also begins to be consumed, although much more slowly than the 3rd layer. These results clearly show that the pseudomorphic, strained 2nd and 3rd layers have higher energy than the 1st Cr layer (which covers the high energy W(110) surface) and the 4th and thicker layers, whose films contain stress-relieving dislocations near the interface with the substrate. A video version is available, Fig25.avi.

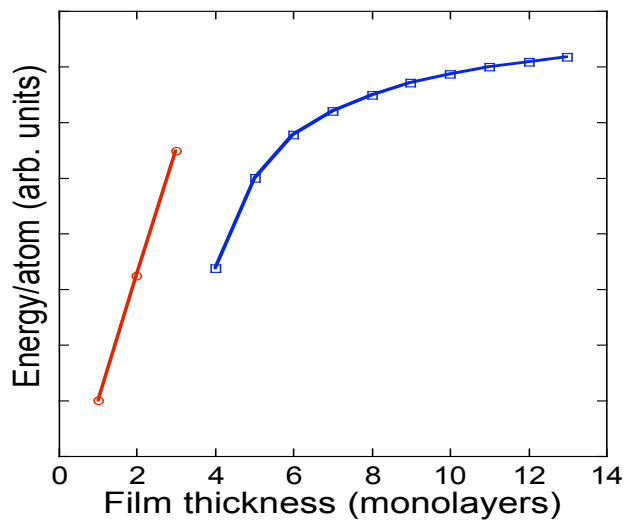

Fig. 26. Schematic illustration of the film energy per atom, referenced to bulk Cr, as a function of thickness. The 1st Cr layer has low energy because bare W has a high surface energy. Strain increases the energy of the 2nd and 3rd layers. Interfacial dislocations relieve strain in films thicker than 3 layers, causing the discontinuity between 3 and 4 layers. The energy dependence of 1-3 layers is sketched and the dependence of more than 3 layers is taken from the EAM simulations of Cu(111) (section 4.4)

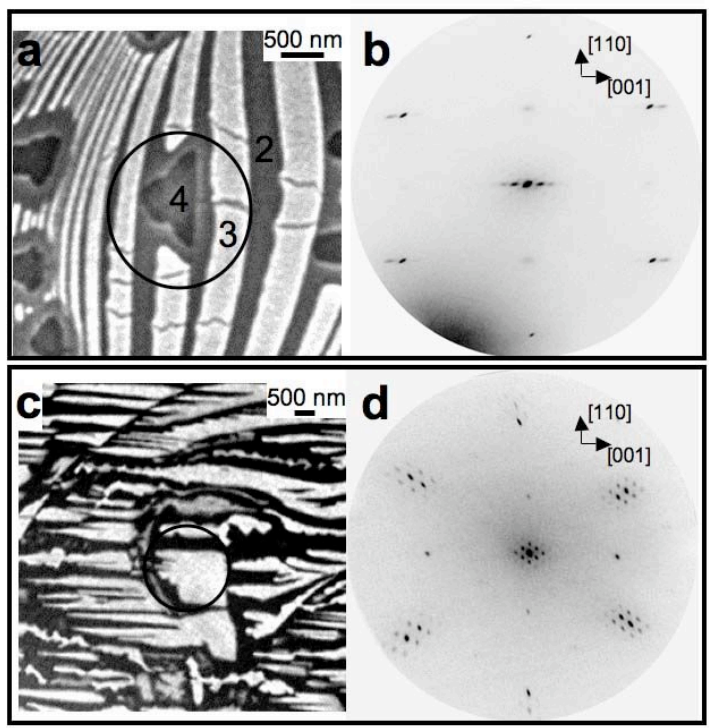

Fig. 27. LEED from two different Cr films that de-wetted during growth at about 350°C. a): LEEM image with layer thicknesses labelled. b). LEED from area within circle of image a (136 eV). The superstructure spots along the [001] direction result from the 4th-layer Cr. c). LEEM image from film of average thickness 10 ML. d). LEED from area within circle of image c (269 eV). The superstructure spots result from the thick (>10 ML) Cr region (bright).